%% file: jfp19.tex
\documentclass{jfp}

\usepackage{microtype}
\usepackage{agda}
\usepackage{catchfilebetweentags}
\usepackage{stmaryrd}
\DisableLigatures[>]{}
\usepackage[T1]{fontenc}
\usepackage{mathpartir}

\input{unicode}
\input{commands}

\let\oldciteyear\citeyear
\renewcommand{\citeyear}[1]{(\oldciteyear{#1})}
\let\oldcite\cite
\renewcommand{\cite}[1]{(\oldcite{#1})}

\newenvironment{agdasnippet}[1][0]
 {\ifx{#1}\undefined
   \centering
  \else
  \fi
}{\ignorespacesafterend%
}

\newcommand{\dquote}[1]{``{#1}''}

\newcommand{\codequote}[1]{\dquote{#1}}
\newcommand{\scarequote}[1]{\dquote{#1}}
\newcommand{\literalquote}[1]{\dquote{#1}}

\begin{document}

\journaltitle{JFP}
\cpr{Cambridge University Press}
\doival{10.1017/xxxxx}

\lefttitle{A type- and scope-safe universe of syntaxes with binding}
\righttitle{Journal of Functional Programming}

\totalpg{\pageref{lastpage01}}
\jnlDoiYr{2021}

\title{A type- and scope-safe universe of syntaxes with binding: their semantics and proofs}

\begin{authgrp}
  \author{Guillaume Allais}\\
  \affiliation{University of St Andrews, St Andrews KY16 9AJ, UK \\
    \email{guillaume.allais@ens-lyon.org}}
  \author{Robert Atkey}
  \affiliation{University of Strathclyde, Glasgow G1 1XQ, UK \\
    \email{robert.atkey@strath.ac.uk}}
  \author{James Chapman}
  \affiliation{Input Output HK Ltd., Edinburgh EH8 9BT, UK \\
    \email{james.chapman@iohk.io}}
  \author{Conor McBride}
  \affiliation{University of Strathclyde, Glasgow G1 1XQ, UK \\
    \email{conor.mcbride@strath.ac.uk}}
  \author{James McKinna}
  \affiliation{Heriot-Watt University, Edinburgh EH14 4AS, UK \\
    \email{J.McKinna@hw.ac.uk}}
\end{authgrp}

\begin{abstract}
The syntax of almost every programming language includes a notion of binder
and corresponding bound occurrences, along with the accompanying
notions of $\alpha$-equivalence, capture-avoiding substitution, typing
contexts, runtime environments, and so on. In the past, implementing
and reasoning about programming languages required careful handling to
maintain the correct behaviour of bound variables. Modern programming
languages include features that enable constraints like scope safety
to be expressed in types. Nevertheless, the programmer is still forced
to write the same boilerplate over again for each new implementation
of a scope-safe operation (e.g., renaming, substitution, desugaring,
printing), and then again for correctness proofs.
We present an expressive universe of syntaxes with binding and
demonstrate how to (1) implement scope-safe traversals once and for
all by generic programming; and (2) how to derive properties of these
traversals by generic proving. Our universe description, generic
traversals and proofs, and our examples have all been formalised in
Agda and are available in the accompanying material available online
at \url{https://github.com/gallais/generic-syntax}.
\end{abstract}

\maketitle

\input{syntax.tex}

\section*{Acknowledgements}

Allais acknowledges the support of EPSRC grant no. EP/M016951/1
\emph{Homotopy Type Theory: Programming and Verification}.
Part of the research leading to these results has also received funding from
the European Research Council under the European Union’s Seventh Framework
Programme (FP7/2007-2013) / ERC grant agreement nr. 320571.

Atkey acknowledges the support of EPSRC grant no. EP/T026960/1
\emph{AISEC: AI Secure and Explainable by Construction}.

Chapman thanks IOHK for their support.

McKinna gratefully acknowledges the support of LFCS, School of
Informatics, University of Edinburgh.

The authors would like to thank the ICFP and JFP reviewers for their
useful comments, and the JFP editors and staff for their patience.

\section*{Conflicts of Interest}

None.

\renewcommand{\bibsection}{}
\section*{References}
\bibliographystyle{jfplike}
\bibliography{strings,main}

\label{lastpage01}

\end{document}

%% file: unicode.tex
\usepackage[utf8]{inputenc}
\usepackage{newunicodechar}
\usepackage{amssymb}
\usepackage{txfonts}

\newunicodechar{⌞}{\ensuremath{\llcorner}}
\newunicodechar{⌟}{\ensuremath{\lrcorner}}
\newunicodechar{′}{\ensuremath{\prime}}
\newunicodechar{−}{\ensuremath{-}}
\newunicodechar{─}{\ensuremath{-}}
\newunicodechar{◆}{\ensuremath{\Diamondblack}}
\newunicodechar{⧫}{\ensuremath{\blacklozenge}}
\newunicodechar{∷}{\ensuremath{::}}
\newunicodechar{∙}{\ensuremath{\bullet}}
\newunicodechar{□}{\ensuremath{\Box}}
\newunicodechar{∎}{\ensuremath{\blacksquare}}
\newunicodechar{⋆}{\ensuremath{\star}}

\newunicodechar{₀}{\ensuremath{_0}}
\newunicodechar{₁}{\ensuremath{_1}}
\newunicodechar{₂}{\ensuremath{_2}}
\newunicodechar{₃}{\ensuremath{_3}}

\newunicodechar{ₑ}{\ensuremath{_e}}
\newunicodechar{ᵢ}{\ensuremath{_i}}
\newunicodechar{ₖ}{\ensuremath{_k}}
\newunicodechar{ₘ}{\ensuremath{_m}}
\newunicodechar{ₙ}{\ensuremath{_n}}
\newunicodechar{ᵣ}{\ensuremath{_r}}
\newunicodechar{ₛ}{\ensuremath{_s}}

\newunicodechar{₊}{\ensuremath{_+}}

\newunicodechar{⁺}{\ensuremath{\textsuperscript{+}}}
\newunicodechar{⁻}{\ensuremath{\textsuperscript{-}}}

\newunicodechar{²}{\ensuremath{^2}}

\newunicodechar{ⁱ}{\ensuremath{^i}}
\newunicodechar{ˡ}{\ensuremath{^l}}
\newunicodechar{ʳ}{\ensuremath{^r}}

\newunicodechar{ᴬ}{\ensuremath{^A}}
\newunicodechar{ᴮ}{\ensuremath{^B}}
\newunicodechar{ᴵ}{\ensuremath{^I}}
\newunicodechar{ᴿ}{\ensuremath{^R}}
\newunicodechar{ᵀ}{\ensuremath{^T}}
\newunicodechar{ᵁ}{\ensuremath{^U}}
\newunicodechar{ⱽ}{\ensuremath{^V}}

\newunicodechar{⋯}{\ensuremath{\cdots}}

\newunicodechar{≡}{\ensuremath{\equiv}}
\newunicodechar{≢}{\ensuremath{\not\equiv}}
\newunicodechar{≟}{\mbox{\tiny\ensuremath{\stackrel{?}{=}}}}
\newunicodechar{≈}{\ensuremath{\approx}}

\newunicodechar{≤}{\ensuremath{\le}}
\newunicodechar{≥}{\ensuremath{\ge}}

\newunicodechar{→}{\ensuremath{\rightarrow}}
\newunicodechar{←}{\ensuremath{\leftarrow}}
\newunicodechar{⇒}{\ensuremath{\Rightarrow}}
\newunicodechar{⇉}{\ensuremath{\rightrightarrows}}

\newunicodechar{∂}{\ensuremath{\partial}}
\newunicodechar{∋}{\ensuremath{\ni}}
\newunicodechar{∞}{\ensuremath{\infty}}
\newunicodechar{∀}{\ensuremath{\forall}}
\newunicodechar{∃}{\ensuremath{\exists}}
\newunicodechar{⊢}{\ensuremath{\vdash}}
\newunicodechar{⟨}{\ensuremath{\langle}}
\newunicodechar{⟩}{\ensuremath{\rangle}}
\newunicodechar{⊤}{\ensuremath{\top}}
\newunicodechar{∘}{\ensuremath{\circ}}
\newunicodechar{⊎}{\ensuremath{\uplus}}
\newunicodechar{×}{\ensuremath{\times}}
\newunicodechar{ℕ}{\ensuremath{\mathbb{N}}}
\newunicodechar{⟦}{\ensuremath{\llbracket}}
\newunicodechar{⟧}{\ensuremath{\rrbracket}}
\newunicodechar{∈}{\ensuremath{\in}}
\newunicodechar{↑}{\ensuremath{\uparrow}}
\newunicodechar{¬}{\ensuremath{\neg}}
\newunicodechar{⊥}{\ensuremath{\bot}}
\newunicodechar{↝}{\ensuremath{\leadsto}}
\newunicodechar{↶}{\ensuremath{\curvearrowleft}}
\newunicodechar{↺}{\ensuremath{\circlearrowleft}}
\newunicodechar{⊔}{\ensuremath{\sqcup}}
\newunicodechar{⨆}{\ensuremath{\bigsqcup}}
\newunicodechar{∩}{\ensuremath{\cap}}
\newunicodechar{∪}{\ensuremath{\cup}}

\newunicodechar{ℓ}{\ensuremath{\ell}}

\newunicodechar{Δ}{\ensuremath{\Delta}}
\newunicodechar{Γ}{\ensuremath{\Gamma}}
\newunicodechar{Σ}{\ensuremath{\Sigma}}
\newunicodechar{Θ}{\ensuremath{\Theta}}
\newunicodechar{Ω}{\ensuremath{\Omega}}

\newunicodechar{α}{\ensuremath{\alpha}}
\newunicodechar{β}{\ensuremath{\beta}}
\newunicodechar{δ}{\ensuremath{\delta}}
\newunicodechar{ε}{\ensuremath{\varepsilon}}
\newunicodechar{φ}{\ensuremath{\phi}}
\newunicodechar{γ}{\ensuremath{\gamma}}
\newunicodechar{ι}{\ensuremath{\iota}}
\newunicodechar{κ}{\ensuremath{\kappa}}
\newunicodechar{λ}{\ensuremath{\lambda}}
\newunicodechar{μ}{\ensuremath{\mu}}
\newunicodechar{ψ}{\ensuremath{\psi}}
\newunicodechar{η}{\ensuremath{\eta}}
\newunicodechar{ρ}{\ensuremath{\rho}}
\newunicodechar{σ}{\ensuremath{\sigma}}
\newunicodechar{τ}{\ensuremath{\tau}}
\newunicodechar{ξ}{\ensuremath{\xi}}
\newunicodechar{ζ}{\ensuremath{\zeta}}
\newunicodechar{Π}{\ensuremath{\Pi}}

\newunicodechar{𝓒}{\ensuremath{\mathcal{C}}}
\newunicodechar{𝓔}{\ensuremath{\mathcal{E}}}
\newunicodechar{𝓕}{\ensuremath{\mathcal{F}}}
\newunicodechar{𝓡}{\ensuremath{\mathcal{R}}}
\newunicodechar{𝓢}{\ensuremath{\mathcal{S}}}
\newunicodechar{𝓣}{\ensuremath{\mathcal{T}}}
\newunicodechar{𝓥}{\ensuremath{\mathcal{V}}}
\newunicodechar{𝓦}{\ensuremath{\mathcal{W}}}

%% file: commands.tex
\newcommand{\AK}{\AgdaKeyword}

\newcommand{\AS}{\AgdaSymbol}
\newcommand{\AStr}{\AgdaString}

\newcommand{\AD}{\AgdaDatatype}
\newcommand{\AF}{\AgdaFunction}
\newcommand{\AR}{\AgdaRecord}
\newcommand{\ARF}{\AgdaField}
\newcommand{\AB}{\AgdaBound}
\newcommand{\AIC}{\AgdaInductiveConstructor}
\newcommand{\AC}{\AgdaComment}

\newcommand{\Set}{\mathbf{Set}}

%% file: syntax.tex
\newcommand{\semrec}{\AR{Semantics}}
\newcommand{\semfun}{\AF{semantics}}

\section{Introduction}

In modern typed programming languages, programmers writing embedded
Domain-Specific Languages (DSLs)~\cite{hudak1996building} and researchers formalising them can now
use the host language's type system to help them. Using Generalised
Algebraic Data Types (GADTs) or the more general indexed families of
Type Theory~\cite{dybjer1994inductive} to represent syntax,
programmers can \emph{statically} enforce some of the invariants in
their languages. For example, managing variable scope is a popular use
case in LEGO, Idris, Coq, Agda and
Haskell~\cite{altenkirch1999monadic,DBLP:conf/gpce/BradyH06,DBLP:journals/jar/HirschowitzM12,DBLP:conf/icfp/KeuchelJ12,BachPoulsen,plfa2018,Eisenberg20}
as directly manipulating raw de Bruijn indices is notoriously
error-prone. Solutions have been proposed that range from enforcing
well-scopedness of variables to ensuring full type correctness. In
short, these techniques use the host languages' types to ensure that
\scarequote{illegal states are unrepresentable}, where illegal states
correspond to ill scoped or ill typed terms in the object language.

Despite the large body of knowledge in how to use types to define well
formed syntax (see the related work in Section
\ref{section:related-work}), it is still necessary for the working DSL
designer or formaliser to redefine essential functions like renaming
and substitution for each new syntax, and then to reprove essential
lemmas about those functions. To reduce the burden of such repeated
work and boilerplate, in this paper we apply the methodology of
data type genericity to programming and proving
  in the domain of
syntaxes with binding.

To motivate our approach, let us look at the formalisation of an
apparently straightforward program transformation: the inlining of
let-bound variables by substitution together with a soundness lemma
proving that reductions in the source languages can be simulated by
reductions in the target one. There are two languages: the source
(\AD{S}), which has let-bindings, and the target (\AD{T}), which only
differs in that it does not:
\begin{displaymath}
  \AD{S} ::= x \mid \AD{S}~\AD{S} \mid \lambda x. \AD{S} \mid \textrm{let }x=\AD{S}\textrm{ in }\AD{S}
  \qquad
  \AD{T} ::= x \mid \AD{T}~\AD{T} \mid \lambda x. \AD{T}
\end{displaymath}

Breaking the task down, an implementer needs to define an operational
semantics for each language, define the program transformation itself,
and prove a correctness lemma that states each step in the source
language is simulated by zero or more steps of the transformed terms
in the target language. In the course of doing this, they will
discover that there is actually a large amount of work:

\begin{enumerate}
\item To define the operational semantics, one needs to define
  substitution, and hence renaming. This needs to be done separately
  for both the source and target languages, even though they are very
  similar;
\item In the course of proving the correctness lemma, one needs to
  prove eight lemmas about the interactions of renaming, substitution,
  and transformation that are all remarkably similar, but must be
  stated and proved separately (e.g, as observed by Benton, Hur, Kennedy
  and McBride~\citeyear{benton2012strongly}).
\end{enumerate}

Even after doing all of this work, they have only a result for a single
pair of source and target languages. If they were to change their
languages $\AD{S}$ or $\AD{T}$, they would have to repeat the same work
all over again (or at least do a lot of cutting, pasting, and
editing).

The main contribution of this paper is that by using the universe of
syntaxes with binding we present in this paper, we are able to solve
this repetition problem \emph{once and for all}.


\paragraph*{Content and Contributions}
To introduce the basic ideas that this paper builds on, we start with
primers on scoped and sorted terms
(Section~\ref{section:primer-term}), scope- and sort-safe programs
acting on them (Section~\ref{section:primer-program}), and
programmable descriptions of data types (Section~\ref{section:data}).
These introductory sections help us build an understanding of the
problem at hand as well as a toolkit that leads us to the novel
content of this paper: a universe of scope-safe syntaxes with binding
(Section~\ref{section:universe}) together with a notion of scope-safe
semantics for these syntaxes (Section~\ref{section:semantics}).  This
gives us the opportunity to write generic implementations of renaming
and substitution (Section~\ref{section:renandsub}), a generic
let-binding removal transformation (generalising the problem stated
above) (Section~\ref{section:letbinding}), and normalisation by
evaluation (Section~\ref{section:nbyeval}). Further, we show how to
construct generic proofs by formally describing what it means for one
semantics to simulate another (Section~\ref{section:simulation}), or
for two semantics to be fusible (Section~\ref{section:fusion}). This
allows us to prove the lemmas required above for renaming,
substitution, and desugaring of let-binders generically, for
\emph{every} syntax in our universe.

\medskip

Our implementation language is
Agda~\cite{norell2009dependently}. However, our techniques are
language independent: any dependently typed language at least as
powerful as Martin-L\"of Type Theory~\cite{martin1982constructive}
equipped with inductive families~\cite{dybjer1994inductive} such as
Coq~\cite{Coq:manual}, Lean~\cite{DBLP:conf/cade/MouraKADR15} or
Idris~\cite{brady2013idris} ought to do.

\medskip

\paragraph*{Changes with respect to the ICFP 2018 version} This paper
is a revised and expanded version of a paper of the same title that
appeared at ICFP 2018. This extended version of the paper includes
many more examples of the use of our universe of syntax with binding
for writing generic programs in Section~\ref{section:catalogue}:
pretty printing with human readable names
(Section~\ref{section:genericprinting}), scope checking
(Section~\ref{section:genericscoping}), type checking
(Section~\ref{section:typechecking}), elaboration
(Section~\ref{section:elaboration}), inlining of single use let-bound
expressions (shrinking reductions) (Section~\ref{section:inlining}),
and normalisation by evaluation (Section~\ref{section:nbyeval}). We
have also included a discussion of how to define generic programs for
deciding equality of terms. Additionally, we have elaborated our
descriptions and examples throughout, and expanded our discussion of
related work in Section~\ref{section:related-work}.


\section{A primer on scope- and sort-safe terms}\label{section:primer-term}

\paragraph*{From inductive types to inductive families for abstract
  syntax.} A reasonable way to represent the abstract syntax of the
untyped $\lambda$-calculus in a typed functional programming language
is to use an inductive type:
\begin{displaymath}
  \begin{array}{l}
    \AK{data}\;\AD{Lam} : \AD{Set}\;\AK{where} \\
    \quad\begin{array}{@{}lll}
           \AIC{`var} &:& \AD{ℕ} \to \AD{Lam} \\
           \AIC{`lam} &:& \AD{Lam} \to \AD{Lam} \\
           \AIC{`app} &:& \AD{Lam} \to \AD{Lam} \to \AD{Lam}
    \end{array}
  \end{array}
\end{displaymath}
We have used de Bruijn~\citeyear{de1972lambda} indices to represent
variables by the number of $\mathsf{`lam}$ binders one has to pass up
through to reach the binding occurrence. The de Bruijn representation
has the advantage that terms are automatically represented up to
$\alpha$-equivalence. If the index goes beyond the number of binders
enclosing it, then we assume that it is referring to some context,
left implicit in this representation.

This representation works well enough for writing programs, but the
programmer must constantly be vigilant to guard against the accidental
construction of ill scoped terms. The implicit context that
accompanies each represented term is prone to being forgotten or
muddled with another, leading to confusing behaviour when variables
either have dangling pointers or point to the wrong thing.

To improve on this situation, previous authors have proposed to use
the host language's type system to make the implicit context explicit,
and to enforce well-scopedness of variables. Scope-safe terms follow
the discipline that every variable is either bound by some binder or
is explicitly accounted for in a context. Bellegarde and
Hook~\citeyear{BELLEGARDE1994287}, Bird and
Paterson~\citeyear{bird_paterson_1999}, and Altenkirch and
Reus~\citeyear{altenkirch1999monadic} introduced the classic
presentation of scope safety using inductive
\emph{families}~\cite{dybjer1994inductive} instead of plain inductive
types to represent abstract syntax. Indeed, using a family indexed by
a $\Set{}$, we can track scoping information at the type level. The
empty $\Set$ represents the empty scope. The type constructor
$1 + (\_)$ extends the running scope with an extra variable.
\begin{displaymath}
  \begin{array}{l}
    \AK{data}\;\AD{Lam} : \AD{Set} \to \AD{Set}\;\AK{where} \\
    \quad\begin{array}{@{}lll}
           \AIC{`var} &:& \AB{X} \to \AD{Lam}\;\AB{X} \\
           \AIC{`lam} &:& \AD{Lam}\;(\AD{1+} \AB{X}) \to \AD{Lam}\;\AB{X} \\
           \AIC{`app} &:& \AD{Lam}\;\AB{X} \to \AD{Lam}\;\AB{X} \to \AD{Lam}\;\AB{X}
    \end{array}
  \end{array}
\end{displaymath}

\paragraph*{Implicit generalisation of variables in Agda.} The careful
reader may have noticed that we use a seemingly out-of-scope variable
\AB{X} of type \AF{Set}. The latest version of Agda allows us to declare
variables that the system should implicitly quantify over if it happens
to find them used in types. This allows us to lighten the presentation
by omitting a large number of prenex quantifiers. The reader will hopefully
be familiar enough with prenex polymorphic types in the style of Standard ML \cite{defnsml} that this will
seem natural to them.

The \AD{Lam} type is now a family of types, indexed by the set
of variables in scope. Thus, the context for each represented term has
been made visible to the type system, and the types enforce that only
variables that have been explicitly declared can be referenced in the
\AIC{`var} constructor. We have made illegal terms
unrepresentable.

Since \AD{Lam} is defined to be a function $\AD{Set} \to \AD{Set}$, it
makes sense to ask whether it is also a functor and a monad. Indeed it
is, as Altenkirch and Reus showed. The functorial action corresponds
to renaming, the monadic \codequote{return} corresponds to the use of variables
(the \AIC{`var} constructor), and the monadic \codequote{bind} corresponds
to substitution. The functor and monad laws correspond to well known
properties from the equational theories of renaming and
substitution. We will revisit these properties, for our whole universe
of syntax with binding, in Section~\ref{section:fusion}.

\paragraph*{A Typed Variant of Altenkirch and Reus' Calculus.}
\label{section:mech-reus}

There is no reason to restrict this technique to inductive families
indexed by $\Set{}$. The more general case of inductive families in
$\Set{}^J$ can be endowed with similar functorial and monadic
operations by using Altenkirch, Chapman and Uustalu's relative
monads~\citeyear{Altenkirch2010, JFR4389}.

We pick as our index type $J$ the category whose objects are
inhabitants of \AD{List} \AB{I} (\AB{I} is a parameter of the
construction) and whose morphisms are \emph{thinnings} (permutations that may
forget elements, we give the definition in Section~\ref{sec:genenvironment}).  Values of type
\AD{List} \AB{I} are intended to represent the list of the sorts (or
kinds, or types, depending on the application) of the de Bruijn
variables in scope. We can recover an unsorted approach by picking $I$
to be the unit type.  Given this sorted setting, our functors take an
extra $I$ argument corresponding to the sort of the expression being
built. This is captured by the large type \AB{I}
\AF{─Scoped}:

\begin{agdasnippet}
\ExecuteMetaData[Data/Var.tex]{scoped}
\end{agdasnippet}
We use Agda's mixfix operator notation, where underscores denote
argument positions.

To lighten the presentation, we exploit the observation that the
current scope is either passed unchanged to subterms (e.g. in the
application case) or extended (e.g. in the λ-abstraction case) by
introducing combinators to build indexed types. We conform to the
convention (see e.g.~\citet{martin1982constructive}) of mentioning
only context \emph{extensions} when presenting judgements.
That is
to say that we aim to write sequents with an \emph{implicit} ambient
context. Concretely: in the simply typed
$\lambda$-calculus (STLC) we would rather use the rule \AB{appᵢ} than
\AB{appₑ} as the inference rule for application.

\begin{mathpar}
\inferrule{f : σ → τ \and t : σ}
 {f\,t : τ}
 {appᵢ}
\and
\inferrule{Γ ⊢ f : σ → τ \and Γ ⊢ t : σ}
 {Γ ⊢ f\,t : τ}
 {appₑ}
\end{mathpar}

In this discipline, the turnstile is used in rules which are binding
fresh variables. It separates the \emph{extension} applied to the ambient
context on its left and the judgment that lives in the thus extended
context on its right. Concretely: we would rather use the rule \AB{lamᵢ}
than \AB{lamₑ} as the inference rule for λ-abstraction in STLC.

\begin{mathpar}
\inferrule{x:σ ⊢ b : τ}
 {λx.t : σ → τ}
 {lamᵢ}
\and
\inferrule{Γ, x:σ ⊢ b : τ}
 {Γ ⊢ λx.t : σ → τ}
 {lamₑ}
\end{mathpar}

This observation that an ambient context is either passed around as is
or extended for subterms is critical to our whole approach to syntax
with binding, and will arise again in our generic formulation of
syntax traversals in Section~\ref{section:semantics}. To facilitate
this, we make use of the following combinators for building indexed
sets:

\noindent
\begin{minipage}{\textwidth}
  \begin{minipage}{0.5\textwidth}
    \ExecuteMetaData[Stdlib.tex]{arrow}
  \end{minipage}\hfill
  \begin{minipage}{0.5\textwidth}
    \ExecuteMetaData[Stdlib.tex]{adjust}
  \end{minipage}
\end{minipage}

\noindent
\begin{minipage}{\textwidth}
  \begin{minipage}{0.5\textwidth}
    \ExecuteMetaData[Stdlib.tex]{constant}
  \end{minipage}\hfill
  \begin{minipage}{0.5\textwidth}
    \ExecuteMetaData[Stdlib.tex]{forall}
  \end{minipage}
\end{minipage}

We lift the function space pointwise with \AF{\_⇒\_}, silently
threading the underlying scope. The \AF{\_⊢\_} makes explicit the
\emph{adjustment} made to the index by a function, a generalisation
of the idea of \emph{extension}. We write \AB{f}
\AF{⊢} \AB{T} where \AB{f} is the adjustment and \AB{T} the indexed
Set it operates on. Although it may seem surprising at first to define
binary infix operators as having arity three, they are meant to be
used partially applied, surrounded by \AF{∀[\_]} which turns an
indexed Set into a Set by implicitly quantifying over the index.
Lastly, \AF{const} is the constant combinator, which ignores the
index.

We make \AF{\_⇒\_} associate to the right as one would expect and give it the
highest precedence level as it is the most used combinator. These combinators
lead to more readable type declarations.  For instance, the compact expression
\AF{∀[} (\AF{const} \AB{P} \AF{⇒} \AIC{s} \AF{⊢} \AB{Q}) \AF{⇒} \AB{R} \AF{]}
desugars to the more verbose type
\AS{∀} \{\AB{i}\} \AS{→} (\AB{P} \AS{→} \AB{Q} (\AIC{s} \AB{i})) \AS{→} \AB{R} \AB{i}.

As the context argument comes second in the definition of
\AF{\_─Scoped}, we can readily use these combinators to thread,
modify, or quantify over the scope when defining such families, as for
example in this data type for scope- and sort-aware de Bruijn indices:

\begin{agdasnippet}
  \ExecuteMetaData[Data/Var.tex]{var}
\end{agdasnippet}

The inductive family \AD{Var} represents well scoped and well sorted
de Bruijn indices. Its \AIC{z} (for zero) constructor refers to the
nearest binder in a non-empty scope. The \AIC{s} (for successor)
constructor lifts a a variable in a given scope to the extended scope
where an extra variable has been bound. Both of the constructors'
types have been written using the combinators defined above.  They
respectively normalise to:
\begin{center}
  \AIC{z} : {\AS{∀} \{\AB{σ} \AB{Γ}\}
            \AS{→} \AD{Var} \AB{σ} (\AB{σ} \AIC{::} \AB{Γ})}
  \qquad
  \AIC{s} : {\AS{∀} \{\AB{σ} \AB{τ} \AB{Γ}\}
            \AS{→} \AD{Var} \AB{σ} \AB{Γ}
            \AS{→} \AD{Var} \AB{σ} (\AB{τ} \AIC{::} \AB{Γ})}
\end{center}
We will reuse the \AD{Var} family to represent variables in all the
syntaxes defined in this paper. We start with the simply typed
$\lambda$-calculus (STLC):

\noindent
\begin{minipage}{0.99\textwidth}
  \begin{minipage}[t]{0.4\textwidth}
    \ExecuteMetaData[StateOfTheArt/ACMM.tex]{type}
  \end{minipage}
  \begin{minipage}[t]{0.59\textwidth}
    \ExecuteMetaData[StateOfTheArt/ACMM.tex]{tm}
  \end{minipage}
\end{minipage}

The \AD{Type} \AF{─Scoped} family \AD{Lam} is Altenkirch and Reus'
intrinsically typed representation of the simply typed λ-calculus,
where \AD{Type} is the Agda type of simple types.  We can readily
write well scoped-and-typed terms such as application, a closed
term of type {((\AB{σ} \AIC{`→} \AB{τ}) \AIC{`→} (\AB{σ} \AIC{`→}
  \AB{τ}))} (\AC{\{-} and \AC{-\}} delimit comments meant to help the
reader see to which binders each de Bruijn index refers):

\begin{agdasnippet}
  \ExecuteMetaData[StateOfTheArt/ACMM.tex]{apply}
\end{agdasnippet}


\section{A primer on type- and scope-safe programs}\label{section:primer-program}

The type- and scope-safe representation described in the previous
section is naturally only a start. Once the programmer has access to a
good representation of the language they are interested in, they will
want to write programs manipulating terms.  Renaming and substitution
are the two typical examples that are required for almost all
syntaxes. Now that well-typedness and well-scopedness are enforced
statically, all of these traversals have to be implemented in a type-
and scope-safe manner.  These constraints show up in the types of
renaming and substitution defined as follows:

\noindent
\begin{minipage}{0.99\textwidth}
\begin{minipage}{0.50\textwidth}
\ExecuteMetaData[StateOfTheArt/ACMM.tex]{ren}
\end{minipage}\hfill
\begin{minipage}{0.49\textwidth}
\ExecuteMetaData[StateOfTheArt/ACMM.tex]{sub}
\end{minipage}
\end{minipage}

We have intentionally hidden technical details behind some auxiliary definitions
left abstract here: \AF{var} and \AF{extend}. Their implementations are distinct
for \AF{ren} and \AF{sub} but they serve the same purpose: \AF{var} is used to
turn a value looked up in the evaluation environment into a term and \AF{extend}
is used to alter the environment when going under a binder. This presentation
highlights the common structure between \AF{ren} and \AF{sub} which we will exploit
later in this section, particularly in Section~\ref{section:lamsem}
where we define an abstract notion of semantics and the corresponding generic traversal.

\subsection{A generic notion of environments}\label{sec:genenvironment}

Both renaming and substitution are defined in terms of \emph{environments}.
We typically call an environment that associates values
to each variable in \AB{Γ} a \AB{Γ}-environment. This informs our notation choice: we write
{((\AB{Γ} \AR{─Env}) \AB{𝓥} \AB{Δ})} for an environment that associates
a value \AB{𝓥} (variables for renaming, terms for substitution) well scoped
and well typed in \AB{Δ} to every entry in \AB{Γ}. Formally, we have the following
record structure (using a record helps Agda's type inference reconstruct the
type family \AB{𝓥} of values for us):

\begin{agdasnippet}
\ExecuteMetaData[Data/Environment.tex]{env}
\end{agdasnippet}

\paragraph*{Environments as records in Agda.}
As with (all) other record structures defined in this paper, we are
able to profit from Agda's \emph{copattern} syntax, as introduced in
\cite{abel2013copatterns} and showcased in \cite{thibodeau2016case}.
That is, when defining an environment \AB{\(\rho\)}, we may either use
the constructor \AIC{pack}, packaging a function \AB{r} as an
environment {\AB{\(\rho\)}~=~\AIC{pack}~\AB{r}}, or else define
\AB{\(\rho\)} in terms of the underlying function obtained from it by
projecting out the (in this case, unique) \ARF{lookup} field, as
{\ARF{lookup}~\AB{\(\rho\)}~=~\AB{r}}. A value of a record type with
more than one field requires each of its fields to be given, either by
a named constructor (or else Agda's default \AK{record} syntax), or in
copattern style. By analogy with record/object syntax in other
languages, Agda further supports \scarequote{dot} notation, so that an equivalent
definition here could be expressed as
{\AB{\(\rho\)}~.\ARF{lookup}~=~\AB{r}}.

We can readily define some basic building blocks for environments:

\noindent
\begin{minipage}{0.99\textwidth}
  \centering
  \begin{minipage}[t]{0.3\textwidth}
    \ExecuteMetaData[Data/Environment.tex]{empty}
  \end{minipage}
  \begin{minipage}[t]{0.69\textwidth}
    \ExecuteMetaData[Data/Environment.tex]{extension}
  \end{minipage}
\end{minipage}
\begin{agdasnippet}
  \ExecuteMetaData[Data/Environment.tex]{envmap}
\end{agdasnippet}
The empty environment (\AF{ε}) is implemented
by remarking that there can be no variable of type
{(\AD{Var} \AB{σ} \AIC{[]})} and to correspondingly dismiss the case with
the impossible pattern \AS{()}. The function \AF{\_∙\_} extends an existing
\AB{Γ}-environment with a new value of type \AB{σ} thus returning a
{(\AB{σ} \AIC{∷} \AB{Γ})}-environment. We also include the definition
of \AF{\_<\$>\_}, which lifts in a pointwise manner a function acting
on values into a function acting on environment of such values.

As we have already observed, the definitions of renaming and substitution have very
similar structure. Abstracting away this shared structure would allow for these
definitions to be refactored, and their common properties to be proved in one swift
move.

Previous efforts in dependently typed
programming~\cite{benton2012strongly,allais2017type}
have achieved this goal and refactored renaming and substitution,
but also normalisation by evaluation, printing with names or continuation-passing style (CPS) conversion
as various instances of a more general traversal. As we will show in Section~\ref{section:typechecking},
type checking in the style of Atkey~\citeyear{atkey2015algebraic} also
fits in that framework. To make sense of this body of work, we
need to introduce three new notions below: \AF{Thinning}, a generalisation of
renaming; the
\AF{□} functor, which freely adds the ability to absorb \AF{Thinning}s to any indexed type; and \AF{Thinnable}s, which are \AF{□}-coalgebras, i.e., types that permit thinning.
We use \AF{□}, and our compact notation for the indexed function space
between indexed types, to crisply encapsulate the additional quantification
over environment extensions which is typical of Kripke semantics.

\paragraph*{The special case of thinnings}~

\begin{agdasnippet}
    \ExecuteMetaData[Data/Environment.tex]{thinning}
\end{agdasnippet}

\AF{Thinning}s subsume more structured notions such as the Category of
Weakenings~\cite{altenkirch1995categorical} or Order Preserving
Embeddings~\cite{chapman2009type}. In particular, they do not prevent the
user from defining arbitrary permutations or from introducing contractions
although we will not use such instances. However, such extra flexibility
will not get in our way, and permits a representation as a function space
which grants us monoid laws \literalquote{for free} as per Jeffrey's
observation~\citeyear{jeffrey2011assoc}. We define the following identity, weaken and (generalised) transitivity combinators for \AF{Thinning}s:

\noindent
\begin{minipage}{0.95\textwidth}
\begin{minipage}{0.45\textwidth}
  \ExecuteMetaData[Data/Environment.tex]{identity}
\end{minipage}
\begin{minipage}{0.45\textwidth}
  \ExecuteMetaData[Data/Environment.tex]{weaken}
\end{minipage}
  \ExecuteMetaData[Data/Environment.tex]{select}
\end{minipage}

Next, the \AF{□} combinator turns any (\AD{List} \AB{I})-indexed Set into one that can
absorb thinnings.

\noindent
\begin{minipage}{0.95\textwidth}
\begin{minipage}{0.45\textwidth}
\ExecuteMetaData[Data/Environment.tex]{box}
\end{minipage}\hfill
\begin{minipage}{0.45\textwidth}
\ExecuteMetaData[Data/Environment.tex]{thinnable}
\end{minipage}

\begin{minipage}{0.26\textwidth}
\ExecuteMetaData[Data/Environment.tex]{extract}
\end{minipage}\hfill
\begin{minipage}{0.37\textwidth}
\ExecuteMetaData[Data/Environment.tex]{duplicate}
\end{minipage}\hfill
\begin{minipage}{0.29\textwidth}
\ExecuteMetaData[Data/Environment.tex]{thBox}
\end{minipage}

\end{minipage}
This is accomplished by abstracting over all possible thinnings
from the current scope, akin to an S4-style necessity modality. The axioms of S4
modal logic incite us to observe that the functor \AF{□} is a comonad: \AF{extract}
applies the identity \AF{Thinning} to its argument, and \AF{duplicate} is obtained
by composing the two \AF{Thinning}s we are given. The expected laws hold trivially
thanks to Jeffrey's trick mentioned above.

The notion of \AF{Thinnable} is the property of being stable under thinnings;
in other words \AF{Thinnable}s are the coalgebras of \AF{□}.
It is a crucial property for values to have if one wants to be able to push
them under binders. From the comonadic structure we get that
the \AF{□} combinator freely turns any (\AD{List} I)-indexed Set into a
\AF{Thinnable} one.

\subsection{A Generic Notion of Semantics}

As we showed in Allais, Chapman, McBride and McKinna
\citeyear{allais2017type}, which we will refer to mnemonically as
ACMM, once equipped with these new notions we can define an abstract
concept of semantics for our type- and scope-safe language. Provided
that a set of constraints on two ({\AF{Type} \AF{─Scoped}}) families
\AB{𝓥} and \AB{𝓒} is satisfied, we will obtain a traversal of the
following type:

\begin{agdasnippet}
\ExecuteMetaData[StateOfTheArt/ACMM.tex]{semtype}
\end{agdasnippet}

Broadly speaking, a semantics turns our deeply embedded abstract syntax
trees into the shallow embedding of the corresponding parametrised higher
order abstract syntax term. We get a choice of useful type- and scope-safe
traversals by using different \scarequote{host languages} for this shallow embedding.

Semantics, specified by a record \semrec{}, are defined in terms
of a choice of values \AB{𝓥} and computations \AB{𝓒}. A semantics must
satisfy constraints on the notions of values \AB{𝓥} and computations \AB{𝓒}
at hand.

In the following paragraphs, we interleave the definition of the record
of constraints \semrec{} with explanations of our choices. It is important
to understand that all of the indented Agda snippets are part of the record's
definition. Some correspond to record fields (highlighted in \ARF{pink})
while others are mere auxiliary definitions (highlighted in \AF{blue}) as
permitted by Agda.

\begin{agdasnippet}
\ExecuteMetaData[StateOfTheArt/ACMM.tex]{rsemtype}\label{section:lamsem}
\end{agdasnippet}
First of all, values \AB{𝓥} should be \AF{Thinnable} so that \semfun{} may push
the environment under binders.
We call this constraint \ARF{th\textasciicircum{}𝓥},
using a caret to generate a mnemonic name: \ARF{th} refers to \emph{th}innable
and \ARF{𝓥} clarifies the family which is proven to be thinnable\footnote{
We use this convention consistently throughout the paper, using names
such as \AF{vl\textasciicircum{}Tm} for the proof that terms are
\AR{VarLike} in Section~\ref{section:semantics}}.

\begin{agdasnippet}
\addtolength{\leftskip}{\parindent}
  \ExecuteMetaData[StateOfTheArt/ACMM.tex]{thV}
\end{agdasnippet}
This constraint allows us to define \AF{extend}, the generalisation of
the two auxiliary definitions we used when defining \AF{ren} and
\AF{sub} at the start of Section~\ref{section:primer-program}, in
terms of the building blocks introduced in
Section~\ref{sec:genenvironment}.  It takes a context extension from
\AB{Δ} to \AB{Θ} in the form of a thinning, an existing evaluation
environment mapping \AB{Γ} variables to \AB{Δ} values and a value
living in the extended context \AB{Θ} and returns an evaluation
environment mapping ({\AB{σ} \AIC{∷} \AB{Γ}}) variables to \AB{Θ}
values.

\begin{agdasnippet}
\addtolength{\leftskip}{\parindent}
\ExecuteMetaData[StateOfTheArt/ACMM.tex]{extend}
\end{agdasnippet}
Second, the set of computations needs to be closed under various
combinators which are the semantical counterparts of the language's
constructors.
For instance in the variable case we obtain a value from the evaluation
environment but we need to return a computation. This means that values
should embed into computations.

\begin{agdasnippet}
\addtolength{\leftskip}{\parindent}
\ExecuteMetaData[StateOfTheArt/ACMM.tex]{var}
\end{agdasnippet}
The semantical counterpart of application is an operation that takes a
representation of a function and a representation of an argument and
produces a representation of the result.

\begin{agdasnippet}
\addtolength{\leftskip}{\parindent}
\ExecuteMetaData[StateOfTheArt/ACMM.tex]{app}
\end{agdasnippet}
The interpretation of the λ-abstraction is of particular interest:
it is a variant on the Kripke function space one can find in normalisation
by evaluation~\cite{berger1991inverse,berger1993program,CoqDybSK,coquand2002formalised}.
In all possible thinnings of the scope at hand, it promises
to deliver a computation whenever it is provided with a value for its newly
bound variable. This is concisely expressed by the constraint's type:

\begin{agdasnippet}
\addtolength{\leftskip}{\parindent}
\ExecuteMetaData[StateOfTheArt/ACMM.tex]{lam}
\end{agdasnippet}


Agda allows us to package the definition of the generic traversal function
\semfun{} together with the fields of the record \semrec{}. This causes the
definition to be specialised and brought into scope for any instance of
\semrec{} the user will define.
We thus realise the promise made earlier, namely that any given
{\semrec{} \AB{𝓥} \AB{𝓒}} induces
a function which, given a value in \AB{𝓥} for each variable in scope,
transforms a  \AD{Lam} term into a computation \AB{𝓒}. This function is the proof of the Fundamental Lemma of Semantics for \AD{Lam}, relative to a given \semrec{} \AB{𝓥} \AB{𝓒}:

\begin{agdasnippet}
  \addtolength{\leftskip}{\parindent}
  \ExecuteMetaData[StateOfTheArt/ACMM.tex]{sem}
\end{agdasnippet}


\subsection{Instances of \texorpdfstring{\AR{Semantics}}{Semantics}}

Recall that each \AR{Semantics} is parametrised by two families: \AB{𝓥}
and \AB{𝓒}. During the evaluation of a term, variables are replaced by
values of type \AB{𝓥} and the overall result is a computation of type \AB{𝓒}.
Coming back to renaming and substitution:

\begin{minipage}{0.95\textwidth}
  \centering
  \begin{minipage}{0.6\textwidth}
    \ExecuteMetaData[StateOfTheArt/ACMM.tex]{semrenfun}
    \ExecuteMetaData[StateOfTheArt/ACMM.tex]{semsubfun}
  \end{minipage}
\end{minipage}

we see that they both fit in the
\semrec{} framework:

\begin{minipage}{0.95\textwidth}
  \begin{minipage}{0.47\textwidth}
    \ExecuteMetaData[StateOfTheArt/ACMM.tex]{semren}
  \end{minipage}\hfill
  \begin{minipage}{0.53\textwidth}
    \ExecuteMetaData[StateOfTheArt/ACMM.tex]{semsub}
  \end{minipage}
\end{minipage}

The family \AB{𝓥} of values is respectively the family
of variables for renaming, and the family of λ-terms for substitution.
In both cases \AB{𝓒} is the family of λ-terms because the result of the
operation will be a term.
We notice that the definition of substitution depends on
the definition of renaming: to be able to push terms under a binder, we need to
have already proven that they are thinnable.
In both cases we use \AF{weaken} defined in Section~\ref{sec:genenvironment} as the definition of the
thinning which embeds \AB{Γ} into {(\AB{σ} \AIC{∷} \AB{Γ})}.

\label{section:printing}
We also include the definition of a basic printer relying on a name
supply to highlight the fact that computations can very well be
effectful. The ability to generate fresh names is given to us by a
monad that here we decide to call \AF{Fresh}. Concretely, \AF{Fresh} is implemented as an instance of the State monad where
the state is a stream of distinct strings:

\begin{agdasnippet}
  \ExecuteMetaData[StateOfTheArt/ACMM.tex]{monad}
\end{agdasnippet}

The \AF{Printing} semantics is defined by using \AF{Name}s (i.e. \AD{String}s)
as values and \AF{Printer}s (i.e. monadic actions in \AF{Fresh} returning a \AD{String})
as computations. We use a \AR{Wrap}per with a type and a context as phantom types
in order to help Agda's inference propagate the appropriate constraints. We define
a function \AF{fresh} that fetches a name from the name supply and makes sure it is
not available anymore.

\begin{agdasnippet}
  \ExecuteMetaData[StateOfTheArt/ACMM.tex]{valprint}
\end{agdasnippet}

\noindent
\begin{minipage}{0.95\textwidth}
\begin{minipage}{0.45\textwidth}
  \ExecuteMetaData[StateOfTheArt/ACMM.tex]{name}
  \ExecuteMetaData[StateOfTheArt/ACMM.tex]{printer}
\end{minipage}
\begin{minipage}{0.5\textwidth}
  \ExecuteMetaData[StateOfTheArt/ACMM.tex]{freshprint}
\end{minipage}
\end{minipage}

The wrapper \AR{Wrap} does not depend on the scope \AB{Γ} so it is
automatically a thinnable functor, that is to say that we have the (used
but not shown here) definitions \AF{map\textasciicircum{}Wrap} witnessing
the functoriality of \AR{Wrap} and \AF{th\textasciicircum{}Wrap} witnessing
its thinnability. We jump straight
to the definition of the printer.

To print a variable, we are handed the \AR{Name} associated to it by the
environment and \AF{return} it immediately.

\begin{agdasnippet}
\ExecuteMetaData[StateOfTheArt/ACMM.tex]{printvar}
\end{agdasnippet}

To print an application, we produce a string representation, \AB{f}, of the term in
function position, then one, \AB{t}, of its argument and combine them by putting the
argument between parentheses.

\begin{agdasnippet}
\ExecuteMetaData[StateOfTheArt/ACMM.tex]{printapp}
\end{agdasnippet}

To print a λ-abstraction, we start by generating a fresh name, \AB{x}, for the
newly bound variable, use that name to generate a string \AB{b} representing the
body of the function to which we prepend a \literalquote{λ} binding the name \AB{x}.

\begin{agdasnippet}
\ExecuteMetaData[StateOfTheArt/ACMM.tex]{printlam}
\end{agdasnippet}

Putting all of these pieces together, we get the \AF{Printing} semantics:

\begin{agdasnippet}
\ExecuteMetaData[StateOfTheArt/ACMM.tex]{semprint}
\end{agdasnippet}

We show how one can use this newly defined semantics to implement \AF{print},
a printer for closed terms assuming that we have already defined \AF{names},
a stream of distinct strings used as our name supply. We show the result of
running \AF{print} on the term \AF{apply}.

\begin{agdasnippet}
  \ExecuteMetaData[StateOfTheArt/ACMM.tex]{print}
  \ExecuteMetaData[StateOfTheArt/ACMM.tex]{applyprint}
\end{agdasnippet}


Both printing and renaming highlight the importance of distinguishing
values and computations: the type of values in their respective
environments is distinct from their type of computations.

All of these examples are already described at length by ACMM~\citeyear{allais2017type}
so we will not spend any
more time on them. In ACMM we have also obtained the simulation and fusion
theorems demonstrating that these traversals are well behaved as
corollaries of more general results expressed in terms of \semfun{}.
We will come back to this in Section~\ref{section:simulation}.

One important observation to make is the tight connection between the constraints
described in \semrec{} and the definition of \AD{Lam}: the semantical counterparts
of the \AD{Lam} constructors are obtained by replacing the recursive occurrences of
the inductive family with either a computation or a Kripke function space depending
on whether an extra variable was bound. This suggests that it ought to be possible
to compute the definition of \semrec{} from the syntax description. Before doing this
in Section~\ref{section:universe}, we need to look at a generic descriptions of
data types.


\section{A primer on universes of data types}\label{section:data}

Chapman, Dagand, McBride and Morris (CDMM)~\citeyear{Chapman:2010:GAL:1863543.1863547}
defined a universe of data types inspired by Dybjer and Setzer's
finite axiomatisation of inductive-recursive definitions~\citeyear{Dybjer1999}
and Benke, Dybjer and Jansson's universes for generic programs and proofs~\citeyear{benke-ugpp}.
This explicit definition of \emph{codes} for data types empowers the
user to write generic programs tackling \emph{all} of the data types
one can obtain this way. In this section we recall the main aspects
of this construction we are interested in to build up our generic
representation of syntaxes with binding.

The first component of the definition of CDMM's universe (defined below) is an inductive type of
\AD{Desc}riptions of strictly positive functors from $\Set{}^J$ to $\Set{}^I$.
These functors correspond to \AB{I}-indexed containers of \AB{J}-indexed
payloads. Keeping these index types distinct prevents mistaking one for the
other when constructing the interpretation of descriptions. Later of course
we can use these containers as the nodes of recursive datastructures by
interpreting some payloads sorts as requests for
subnodes~\cite{DBLP:journals/jfp/AltenkirchGHMM15}.

The inductive type of descriptions has three constructors:
\AIC{`σ} to store data (the rest of
the description can depend upon this stored value), \AIC{`X} to attach a
recursive substructure indexed by $J$ and \AIC{`$\blacksquare$} to stop
with a particular index value.

The recursive function \AF{⟦\_⟧} makes the interpretation of the
descriptions formal. Interpretation of descriptions give rise to
right-nested tuples terminated by equality constraints.

\noindent
\begin{minipage}{\textwidth}
  \begin{minipage}{0.52\textwidth}
    \noindent
    \ExecuteMetaData[StateOfTheArt/CDMM.tex]{desc}
  \end{minipage}
  \begin{minipage}{0.47\textwidth}
    \ExecuteMetaData[StateOfTheArt/CDMM.tex]{interp}
  \end{minipage}
\end{minipage}

These constructors give the programmer the ability to build up the data
types they are used to. For instance, the functor corresponding
to lists of elements in $A$ stores a \AD{Bool}ean which stands for whether
the current node is the empty list or not. Depending on its value, the
rest of the description is either the \codequote{stop} token or a pair of an element
in $A$ and a recursive substructure, that is, the tail of the list. The \AD{List} type
is unindexed, and we represent the lack of an index with the unit type \AD{$\top$}
whose unique inhabitant is \AIC{tt}.

\begin{agdasnippet}
\ExecuteMetaData[StateOfTheArt/CDMM.tex]{listD}
\end{agdasnippet}

Indices can be used to enforce invariants. For example, the type {\AD{Vec} \AB{A} \AB{n}}
of length-indexed lists. It has the same structure as the definition of \AF{listD}.
We start with a \AF{Bool}ean distinguishing the two constructors: either
the empty list (in which case the branch's index is enforced to be $0$) or a
non-empty one in which case we store a natural number \AB{n}, the head of type
\AB{A} and a tail of size \AB{n} and the branch's index is enforced to be
\AIC{suc} \AB{n}.

\begin{agdasnippet}
\ExecuteMetaData[StateOfTheArt/CDMM.tex]{vecD}
\end{agdasnippet}

The pay-off for encoding our data types as descriptions is that we can define
generic programs for whole classes of data types. The decoding function \AF{⟦\_⟧}
acted on the objects of $\Set{}^J$, and we will now define the function \AF{fmap} by
recursion over a code \AB{d}. It describes the action of the functor corresponding
to \AB{d} over morphisms in $\Set{}^J$. This is the first example of generic
programming over all the functors one can obtain as the meaning of a description.

\begin{agdasnippet}
\ExecuteMetaData[StateOfTheArt/CDMM.tex]{fmap}
\end{agdasnippet}

All the functors obtained as meanings of \AD{Desc}riptions are strictly
positive. So we can build the least fixpoint of the ones that are endofunctors
(i.e. the ones for which $I$ equals $J$). This fixpoint is called \AD{μ}
and its iterator is given by the definition of \AF{fold} \AB{d}
\footnote{The \AD{Size}~\cite{DBLP:journals/corr/abs-1012-4896} index added
to the inductive definition of \AD{μ} plays a crucial role in getting the
termination checker to see that \AF{fold} is a total function.
}
.

\begin{agdasnippet}
\ExecuteMetaData[StateOfTheArt/CDMM.tex]{mu}
\ExecuteMetaData[StateOfTheArt/CDMM.tex]{fold}
\end{agdasnippet}

This least fixpoint allows us to recover the data types we would
otherwise declare recursively and generatively. Pattern synonyms let us hide away the
encoding: programmers can use them to pattern-match on lists and Agda
conveniently resugars them when displaying a goal. Finally, we can get
our hands on the types' eliminators by instantiating the generic
\AF{fold}:

\noindent
\begin{minipage}{0.95\textwidth}
\begin{minipage}[t]{0.48\textwidth}
  \ExecuteMetaData[StateOfTheArt/CDMM.tex]{list}
  \ExecuteMetaData[StateOfTheArt/CDMM.tex]{nilcons}
\end{minipage}
\begin{minipage}[t]{0.50\textwidth}
  \ExecuteMetaData[StateOfTheArt/CDMM.tex]{foldr}
\end{minipage}
\end{minipage}

The CDMM approach, therefore, allows us to generically define iteration principles
for all data types that can be described. These are exactly the features we desire
for a universe of data types with binding, so in the next section we will see how
to extend CDMM's approach to include binding.

The functor underlying any well scoped and sorted syntax can be coded as some
{\AD{Desc} (\AB{I} \AR{×} \AD{List} \AB{I}) (\AB{I} \AR{×} \AD{List} \AB{I})},
 with the
free monad construction from CDMM uniformly adding the variable case. While a
good start, \AD{Desc} treats its index types as unstructured, so this construction
is blind to what makes the {\AD{List} \AB{I}} index a \emph{scope}.
The resulting
\codequote{bind} operator demands a function which maps variables in \emph{any} sort and
scope to terms in the \emph{same} sort and scope. However, the behaviour we need
is to preserve sort while mapping between specific source and target scopes which
may differ. We need to account for the fact that scopes change only by extension,
and hence that our specifically scoped operations can be pushed under binders by
weakening.


\section{A universe of scope-safe and well sorted syntaxes}\label{section:universe}

Our universe of scope-safe and well sorted syntaxes follows the same principle
as CDMM's universe of data types, except that we are not building endofunctors on
$\Set{}^I$ any more but rather on {\AB{I} \AF{─Scoped}}. We now think of the
index type \AB{I} as the sorts
used to distinguish terms in our embedded language. The \AIC{`$\sigma$} and
\AIC{`∎} constructors are as in the CDMM \AD{Desc} type and are used to
represent data and index constraints respectively.
What distinguishes this new universe \AD{Desc} from that of Section~\ref{section:data}
is that the
\AIC{`X} constructor
is now augmented with an additional {\AD{List} \AB{I}} argument that describes
the new binders that are brought into scope at this recursive position. This
list of the sorts of the newly bound variables will play a crucial role when
defining the description's semantics as a binding structure below. 

\begin{agdasnippet}
\ExecuteMetaData[Generic/Syntax.tex]{desc}
\end{agdasnippet}

The meaning function \AF{⟦\_⟧} we associate to a description follows closely
its CDMM equivalent. It only departs from it in the \AIC{`X} case and the fact
it is not an endofunctor on \AB{I} \AF{─Scoped}; it is more general than that.
The function takes an \AB{X} of type {\AD{List} \AB{I} $\rightarrow$ \AB{I} \AD{─Scoped}}
to interpret {\AIC{`X} \AB{Δ} \AB{j}} (i.e. substructures of sort \AB{j} with
newly bound variables in \AB{Δ}) in an ambient scope \AB{Γ} as {\AB{X} \AB{Δ} \AB{j} \AB{Γ}}.

\begin{agdasnippet}
\ExecuteMetaData[Generic/Syntax.tex]{interp}
\end{agdasnippet}

The astute reader may have noticed that \AF{⟦\_⟧} is uniform in $X$ and $\Gamma$; however
refactoring \AF{⟦\_⟧} to use the partially applied $X\,\_\,\_\,\Gamma$ following
this observation would lead to a definition harder to use with the
combinators for indexed sets described in Section \ref{section:mech-reus}
which make our types much more readable.

If we pre-compose the meaning function \AF{⟦\_⟧} with a notion of \scarequote{de Bruijn scopes}
(denoted \AF{Scope} here) which turns any \AB{I} \AF{─Scoped} family into a function
of type \AD{List} \AB{I} \AS{→} \AB{I} \AF{─Scoped} by appending the two
\AD{List} indices, we recover a meaning function producing an endofunctor on
\AB{I} \AF{─Scoped}. So far we have only shown the action of the functor on objects;
its action on morphisms is given by a function \AF{fmap} defined by induction over
the description just as in Section~\ref{section:data}.

\begin{agdasnippet}
\ExecuteMetaData[Generic/Syntax.tex]{scope}
\end{agdasnippet}

The endofunctors thus defined are strictly positive and we can take their fixpoints.
As we want to define the terms of a language with variables, instead of
considering the initial algebra, this time we opt for the free relative
monad~\cite{JFR4389} (with respect to the functor \AF{Var}): the \AIC{`var}
constructor corresponds to return, and we will define bind (also known as
the parallel substitution \AF{sub}) in the next section.


\begin{agdasnippet}
\ExecuteMetaData[Generic/Syntax.tex]{mu}
\end{agdasnippet}

Coming back to our original examples, we now have the ability to give
codes for the well scoped untyped λ-calculus and, just as well,
the intrinsically typed STLC. We add a third
example to showcase the whole spectrum of syntaxes: a well scoped and
well sorted but not well typed bidirectional language. In all examples,
the variable case will be added by the free monad construction so we only
have to describe the other constructors.

\paragraph*{Un(i)typed λ-calculus (UTLC).} For the untyped case, the lack of
type translates to picking the unit type (\AR{⊤}) as our notion of sort.
We have two possible
constructors: application where we have two substructures which do not bind
any extra argument and λ-abstraction which has exactly one substructure
with precisely one extra bound variable. A single \AD{Bool}ean is enough to
distinguish the two constructors.

\begin{agdasnippet}
  \ExecuteMetaData[Generic/Syntax/UTLC.tex]{ULC}
\end{agdasnippet}

\paragraph*{Bidirectional STLC.}\label{par:bidirectional} Our second example is a
bidirectional~\cite{pierce2000local} language hence the introduction of a
notion of \AD{Mode}: each term is either part of the \AIC{Infer} or
\AIC{Check} fraction of the language. This language has four constructors
which we list in the ad hoc \AD{`Bidi} type of constructor tags, its
decoding \AD{Bidi} is defined by a pattern-matching λ-expression in Agda.
Application and λ-abstraction behave as expected, with the important
observation that λ-abstraction binds an \AIC{Infer}rable term. The two
remaining constructors correspond to changes of direction: one can freely
\AIC{Emb}bed inferrable terms as checkable ones whereas we require a type
annotation when forming a \AIC{Cut} (we reuse the notion of \AD{Type} introduced
in the STLC example at the end of Section~\ref{section:primer-term}). 

\noindent
\begin{minipage}{0.95\textwidth}
\begin{minipage}[t]{0.3\textwidth}
  \ExecuteMetaData[Generic/Syntax/Bidirectional.tex]{tagmode}
\end{minipage}\quad
\begin{minipage}[t]{0.65\textwidth}
  \ExecuteMetaData[Generic/Syntax/Bidirectional.tex]{desc}
\end{minipage}
\end{minipage}

\paragraph*{Intrinsically typed STLC.}\label{par:intrinsicSTLC}
In the typed case (for the same notion of \AD{Type}), we are back to two
constructors: the terms are fully annotated and therefore it is not necessary
to distinguish between \AD{Mode}s anymore. We need our tags to carry extra
information about the types involved so we use once more an ad hoc data type
\AD{`STLC}, and define its decoding \AD{STLC} by a pattern-matching λ-expression.

\noindent
\begin{minipage}{0.95\textwidth}
\begin{minipage}[t]{0.48\textwidth}
  \ExecuteMetaData[Generic/Syntax/STLC.tex]{tag}
\end{minipage}
\begin{minipage}[t]{0.515\textwidth}
  \ExecuteMetaData[Generic/Syntax/STLC.tex]{desc}
\end{minipage}
\end{minipage}

For convenience we use Agda's pattern synonyms corresponding to the
original constructors in Section \ref{section:mech-reus}. These
synonyms can be used when pattern-matching on a term and Agda resugars
them when displaying a goal. This means that the end user can
seamlessly work with encoded terms without dealing with the gnarly
details of the encoding.  These pattern definitions can omit some
arguments using \codequote{\AS{\_}}, in which case they will be filled in
by unification just like any other implicit argument: there is no
extra cost to using an encoding!  The only downside is that the
language currently does not allow the user to specify type annotations
for pattern synonyms. We only include examples of pattern synonyms
for the two extreme examples, the definition for \AF{Bidi} are similar.

\noindent
\begin{minipage}{0.99\textwidth}
\begin{minipage}{0.47\textwidth}
  \ExecuteMetaData[Generic/Syntax/UTLC.tex]{LCpat}
\end{minipage}
\begin{minipage}{0.52\textwidth}
  \ExecuteMetaData[Generic/Syntax/STLC.tex]{patST}
\end{minipage}
\end{minipage}

As a usage example of these pattern synonyms, we define the identity
function in all three languages, using the
same caret-based naming convention we introduced earlier. The code
is virtually the same except for \AF{Bidi} which explicitly records
the change of direction from \AIC{Check} to \AIC{Infer}.

\noindent
\begin{minipage}{\textwidth}
\begin{minipage}{0.28\textwidth}
  \ExecuteMetaData[Generic/Syntax/UTLC.tex]{LCid}
\end{minipage}
\begin{minipage}{0.34\textwidth}
  \ExecuteMetaData[Generic/Syntax/Bidirectional.tex]{BDid}
\end{minipage}
\begin{minipage}{0.36\textwidth}
  \ExecuteMetaData[Generic/Syntax/STLC.tex]{STid}
\end{minipage}
\end{minipage}

\paragraph*{A sum combinator for syntaxes.}\label{desccomb}

The definition of \AF{UTLC} is the third time (the first and second
times being the definition of \AF{listD} and \AF{vecD} in
Section~\ref{section:data}) that we use a \AF{Bool} to distinguish
between two constructors. We can abstract this common pattern as a combinator \AF{\_`+\_} together
with an appropriate eliminator \AF{case} which, given two methods,
picks the one corresponding to the chosen branch.

\noindent
\begin{minipage}{0.95\textwidth}
\begin{minipage}[t]{0.42\textwidth}
  \ExecuteMetaData[Generic/Syntax.tex]{descsum}
\end{minipage}
\begin{minipage}[t]{0.57\textwidth}
  \ExecuteMetaData[Generic/Syntax.tex]{case}
\end{minipage}
\end{minipage}



A concrete use case for this combinator will be given in
Section~\ref{section:letbinding}
where we explain how to seamlessly enrich an existing syntax with let-bindings
and how to use the \semrec{} framework to elaborate them away.


\section{Generic scope-safe and well sorted programs for syntaxes}\label{section:semantics}

Based on the \semrec{} type we defined for the specific example of the
simply typed λ-calculus in Section~\ref{section:primer-program},
we can define a generic notion of
semantics for all syntax descriptions. It is once more parametrised
by two \AB{I}\AF{─Scoped} families \AB{𝓥} and \AB{𝓒} corresponding,
respectively, to \emph{values} associated to bound variables and
\emph{computations} delivered by evaluating terms. These two families
have to abide by three constraints:
\begin{itemize}
\item{\ARF{th\textasciicircum{}𝓥}} Values should be thinnable so that we can push the
      evaluation environment under binders;
\item{\ARF{var}} Values should embed into computations for us to be able
      to return the value associated to a variable as the
      result of its evaluation;
\item{\ARF{alg}} We should have an algebra turning
      a term whose substructures have been replaced with
      computations (possibly under some binders, represented semantically
      by the \AF{Kripke} type-valued function defined below) into computations
\end{itemize}

\begin{agdasnippet}
\ExecuteMetaData[Generic/Semantics.tex]{semantics}
\end{agdasnippet}

Here we crucially use the fact that the meaning of a description is
defined in terms of a function interpreting substructures which has
the type \AF{List} \AB{I} \AS{→} \AB{I}\AF{─Scoped}, that is, that gets access
to the current scope but also the exact list of the sorts of the newly bound variables.
We define a function \AF{Kripke} by case analysis on the number of newly bound
variables. It is essentially a subcomputation waiting for a value associated to
each one of the fresh variables.
\begin{itemize}
\item If it is $0$ we expect the substructure to be a computation corresponding
    to the result of the evaluation function's recursive call;
  \item But if there are newly bound variables then we expect to have a function
    space. In any context extension, it will take an environment of values for
    the newly bound variables and produce a computation corresponding to the
    evaluation of the body of the binder.
\end{itemize}

\begin{agdasnippet}
\ExecuteMetaData[Data/Environment.tex]{kripke}
\end{agdasnippet}

It is once more the case that the abstract notion of Semantics comes
with a fundamental lemma: all \AB{I} \AF{─Scoped} families \AB{𝓥} and
\AB{𝓒} satisfying the three criteria we have put forward give rise
to an evaluation function. We introduce a notion of computation
\AF{\_─Comp} analogous to that of environments: instead of associating
values to variables, it associates computations to terms.

\begin{agdasnippet}
  \ExecuteMetaData[Generic/Semantics.tex]{comp}
\end{agdasnippet}

\subsection{Fundamental lemma of semantics}\label{sec:fundamentallemma}

We can now define the type of the fundamental lemma (called \semfun{}) which
takes a semantics and returns a function from environments to computations.
It is defined mutually with a
function \AF{body} turning syntactic binders into semantic binders: to
each de Bruijn \AF{Scope} (i.e. a substructure in a potentially extended
context) it associates a \AF{Kripke} (i.e. a subcomputation expecting a
value for each newly bound variable).

\begin{agdasnippet}
  \ExecuteMetaData[Generic/Semantics.tex]{semtype}
\end{agdasnippet}

The \semfun{} proof is straightforward now that we have clearly
identified the problem structure and the constraints we need to enforce.
If the term considered is a variable, we look up the associated value in
the evaluation environment and turn it into a computation using \ARF{var}.
If it is a non-variable constructor then we call \AF{fmap} to evaluate the
substructures using \AF{body} and then call the \ARF{alg}ebra to combine
these results.

\begin{agdasnippet}
  \ExecuteMetaData[Generic/Semantics.tex]{semproof}
\end{agdasnippet}

The auxiliary lemma \AF{body} distinguishes two cases. If no new
variable has been bound in the recursive substructure, it is
a matter of calling \semfun{} recursively. Otherwise we are provided
with a \AF{Thinning}, some additional values and evaluate the
substructure in the thinned and extended evaluation environment
(thanks to a auxiliary function \AF{\_>>\_} which given two environments
{(\AB{Γ} \AR{─Env}) \AB{𝓥} \AB{Θ}} and {(\AB{Δ} \AR{─Env}) \AB{𝓥} \AB{Θ}}
produces an environment {((\AB{Γ} \AF{++} \AB{Δ}) \AR{─Env}) \AB{𝓥} \AB{Θ})}.

\begin{agdasnippet}
  \ExecuteMetaData[Generic/Semantics.tex]{bodyproof}
\end{agdasnippet}

Given that \AF{fmap} introduces one level of indirection between the
recursive calls and the subterms they are acting upon, the fact
that our terms are indexed by a \AF{Size} is once more crucial in
getting the termination checker to see that our proof is indeed
well founded.

We immediately introduce \AF{closed}, a corollary of the fundamental lemma of
semantics for the special cases of closed terms.

\begin{agdasnippet}
  \ExecuteMetaData[Generic/Semantics.tex]{closed}
\end{agdasnippet}

Given a \AR{Semantics} with value type \AB{𝓥} and computation type \AB{𝓒},
we can evaluate a closed term of type \AB{σ} and obtain a computation of
type {(\AB{𝓒} \AB{σ} \AIC{[]})} by kickstarting the evaluation with an
empty environment.

\subsection{Our first generic programs: renaming and substitution}%
\label{section:renandsub}

Similarly to ACMM~\citeyear{allais2017type} renaming can be defined generically
for all syntax descriptions as a semantics with \AF{Var} as values and \AD{Tm} as
computations. The first two constraints on \AF{Var} described earlier are trivially
satisfied. Observing that renaming strictly respects the structure of the term it
goes through, it makes sense for the algebra to be implemented using \AF{fmap}.
When dealing with the body of a binder, we \scarequote{reify} the \AF{Kripke} function by
evaluating it in an extended context and feeding it placeholder values corresponding to
the extra variables introduced by that context. This is reminiscent both of what we
did in Section~\ref{section:primer-program} and the definition of reification in
the setting of normalisation by evaluation
(see e.g. Catarina Coquand's formal development~\citeyear{coquand2002formalised}).

Substitution is defined in a similar manner with \AD{Tm} as both
values and computations. Of the two constraints applying to terms as
values, the first one corresponds to renaming and the second
one is trivial. The algebra is once more defined by using \AF{fmap}
and reifying the bodies of binders.

\noindent
\begin{minipage}{\textwidth}
\begin{minipage}{0.5\textwidth}
  \ExecuteMetaData[Generic/Semantics/Syntactic.tex]{renaming}
  \ExecuteMetaData[Generic/Semantics/Syntactic.tex]{ren}
\end{minipage}\hfill
\begin{minipage}{0.5\textwidth}
  \ExecuteMetaData[Generic/Semantics/Syntactic.tex]{substitution}
  \ExecuteMetaData[Generic/Semantics/Syntactic.tex]{sub}
\end{minipage}
\end{minipage}

The reification process mentioned in the definition of renaming and
substitution can be implemented generically for \semrec{} families
which have \AR{VarLike} values, that is, values which are \AF{Thinnable} and
such that we can craft placeholder values in non-empty contexts. It is
almost immediate that both \AD{Var} and \AD{Tm} are \AR{VarLike} (with
proofs \AF{vl\textasciicircum{}Var} and \AF{vl\textasciicircum{}Tm},
respectively).

\begin{agdasnippet}
  \ExecuteMetaData[Data/Var/Varlike.tex]{varlike}
\end{agdasnippet}

\label{sec:varlike:base}
Given a proof that \AB{𝓥} is \AR{VarLike}, we can manufacture
  several useful environments of values \AB{𝓥}. We provide users with
  \AF{base} of type {(\AB{Γ} \AR{─Env}) \AB{𝓥} \AB{Γ}},
  \AF{fresh\textsuperscript{r}} of type {(\AB{Γ} \AR{─Env}) \AB{𝓥} (\AB{Δ} \AF{++} \AB{Γ})}
  and \AF{fresh\textsuperscript{l}} of type {(\AB{Γ} \AR{─Env}) \AB{𝓥} (\AB{Γ} \AF{++} \AB{Δ})}
  by combining the use of placeholder values and thinnings.
  In the \AD{Var} case these very general definitions respectively specialise
  to the identity renaming for a context \AB{Γ} and the injection of \AB{Γ}
  fresh variables to the right or the left of an ambient context \AB{Δ}.
  Similarly, in the \AD{Tm} case, we can show \AF{base} \AF{vl\textasciicircum{}Tm}
  extensionally equal to the identity environment \AF{id\textasciicircum{}Tm}
  given by {\ARF{lookup} \AF{id\textasciicircum{}Tm} = \AIC{`var}},
  which associates each variable to itself (seen as a term).
Using these definitions, we can then implement \AF{reify} as follows:

\begin{agdasnippet}
  \ExecuteMetaData[Data/Var/Varlike.tex]{reify}
\end{agdasnippet}


\section{A catalogue of generic programs for syntax with binding}
\label{section:catalogue}

In this section we explore a large part of the spectrum of traversals a
compiler writer may need when implementing their own language.
In Section~\ref{section:genericprinting} we look at the production of
human-readable representations of internal syntax; in Section~\ref{section:genericscoping}
we write a generic scope checker thus bridging the gap between raw data
fresh out of a parser to well scoped syntax; we then demonstrate how to
write a type checker in Section~\ref{section:typechecking} and even an
elaboration function turning well scoped into well scoped and typed syntax
in Section~\ref{section:elaboration}. We then study type and scope respecting
transformations on internal syntax: desugaring in Section~\ref{section:letbinding}
and size preserving inlining in Section~\ref{section:inlining}. We conclude
with an unsafe but generic evaluator defined using normalisation by evaluation
in Section~\ref{section:nbyeval}.

\input{catalogue/printing.tex}
\input{catalogue/scopechecking.tex}
\input{catalogue/typechecking.tex}
\input{catalogue/elaborating.tex}
\input{catalogue/desugaring.tex}
\input{catalogue/inlining.tex}
\input{catalogue/normalising.tex}


\section{Other opportunities for generic programming}

Some generic programs of interest do not fit in the \AR{Semantics}
framework. They can still be implemented once and for all, and even
benefit from the \AR{Semantics}-based definitions.

We will first explore existing work on representing cyclic structures
using a syntax with binding: a binder is a tree node declaring a pointer
giving subtrees the ability to point back to it, thus forming a cycle.
Substitution will naturally play a central role in giving these finite
terms a semantics as their potentially infinite unfolding.

We will then see that many of the standard traversals produced by the
\codequote{deriving} machinery familiar to Haskell programmers can be implemented
on syntaxes too, sometimes with more informative types.

\input{catalogue/unrolling.tex}
\input{catalogue/equality.tex}


\section{Building generic proofs about generic programs}

In ACMM~\citeyear{allais2017type} we have
already shown that, for the simply typed λ-calculus, introducing an abstract
notion of \AF{Semantics} not only reveals the shared structure of common
traversals, it also allows us to give abstract proof frameworks for
simulation or fusion lemmas. This idea naturally extends to our generic
presentation of semantics for all syntaxes.

\subsection{Relations and relation transformers}

In our exploration of generic proofs about the behaviour of various \AR{Semantics},
we are going to need to manipulate relations between distinct notions of values or
computations. In this section, we introduce the notion of relation we are going to
use as well as these two key relation transformers.

In Section~\ref{sec:genenvironment} we introduced a generic notion of well typed
and scoped environment as a function from variables to values. Its formal definition
is given as a record type. This record wrapper helps
Agda's type inference reconstruct the type family of values whenever it is passed an
environment.

For the same reason, we will use a record wrapper for the concrete implementation of
our notion of relation over (I \AF{─Scoped}) families. A \AR{Rel}ation between two
such families \AB{T} and \AB{U} is a function which to any \AB{σ} and \AB{Γ} associates
a relation between (\AB{T} \AB{σ} \AB{Γ}) and (\AB{U} \AB{σ} \AB{Γ}). Our first example
of such a relation is \AF{Eqᴿ} the equality relation between an (\AB{I}\AF{─Scoped})
family \AB{T} and itself.

\noindent
\begin{minipage}{\textwidth}
  \begin{minipage}[t]{0.65\textwidth}
    \ExecuteMetaData[Data/Relation.tex]{rel}
  \end{minipage}
  \begin{minipage}[t]{0.25\textwidth}
    \ExecuteMetaData[Data/Relation.tex]{eqR}
  \end{minipage}
\end{minipage}

Once we know what relations are, we are going to have to lift relations on values
and computations to relations on environments, \AF{Kripke} function spaces or
on \AB{d}-shaped terms whose subterms have been evaluated already.
This is what the rest of this section focuses on.

\paragraph*{Environment relator.}
Provided a relation \AB{𝓥ᴿ} for notions of values \AB{𝓥ᴬ} and \AB{𝓥ᴮ}, by
pointwise lifting we can define a relation {(\AR{All} \AB{𝓥ᴿ} \AB{Γ})} on
\AB{Γ}-environments of values \AB{𝓥ᴬ} and \AB{𝓥ᴮ} respectively. We once more
use a record wrapper simply to facilitate Agda's job when reconstructing
implicit arguments.

\begin{agdasnippet}
  \ExecuteMetaData[Data/Relation.tex]{all}
\end{agdasnippet}

The first example of two environment being related is \AF{reflᴿ} that, to any
environment \AB{ρ} associates a trivial proof of the statement
{(\AR{All} \AF{Eqᴿ} \AB{Γ} \AB{ρ} \AB{ρ})}.
The combinators we introduced in Section~\ref{sec:genenvironment} to build environments
(\AF{ε}, \AF{\_∙\_}, etc.) have natural relational counterparts. We reuse the same
names for them, simply appending an \AF{ᴿ} suffix.

\paragraph*{Kripke relator.}
We assume that we have two types of values \AB{𝓥ᴬ} and \AB{𝓥ᴮ}
as well as a relation \AB{𝓥ᴿ} for pairs of such values, and two types of computations
\AB{𝓒ᴬ} and \AB{𝓒ᴮ} whose notion of relatedness is given by \AB{𝓒ᴿ}. We can define
\AF{Kripkeᴿ} relating Kripke functions of type
{(\AF{Kripke} \AB{𝓥ᴬ} \AB{𝓒ᴬ})} and {(\AF{Kripke} \AB{𝓥ᴮ} \AB{𝓒ᴮ})}
respectively by stating that they send related inputs
to related outputs. We use the relation transformer \AF{All} defined in the previous
paragraph.

\begin{agdasnippet}
  \ExecuteMetaData[Data/Var/Varlike.tex]{kripkeR}
\end{agdasnippet}

\paragraph*{Desc relator.}
The relator (\AF{⟦} \AB{d} \AF{⟧ᴿ}) is a relation transformer which characterises
structurally equal layers such that their substructures are themselves related
by the relation it is passed as an argument. It inherits a lot of its relational
arguments' properties: whenever \AB{R} is reflexive (respectively symmetric or
transitive) so is {(\AF{⟦} \AB{d} \AF{⟧ᴿ} \AB{R})}.\label{lem:zipstable}

It is defined by induction on the description and case analysis on the two
layers which are meant to be equal:
\begin{itemize}
  \item In the stop token case \AIC{`∎} \AB{i}, the two layers are considered to
    be trivially equal (i.e. the constraint generated is the unit type)
  \item When facing a recursive position {\AIC{`X} \AB{$\Delta$} \AB{j} \AB{d}}, we
    demand that the two substructures are related by {\AB{R} \AB{$\Delta$} \AB{j}}
    and that the rest of the layers are related by (\AF{⟦} \AB{d} \AF{⟧ᴿ} \AB{R})
  \item Two nodes of type {\AIC{`$\sigma$} \AB{A} \AB{d}} will
    be related if they both carry the same payload \AB{a} of type \AB{A} and if
    the rest of the layers are related by (\AF{⟦} \AB{d} \AB{a} \AF{⟧ᴿ} \AB{R})
\end{itemize}

\begin{agdasnippet}
  \ExecuteMetaData[Generic/Relator.tex]{ziptype}
\end{agdasnippet}

If we were to take a fixpoint of \AF{⟦\_⟧ᴿ}, we could obtain a structural
notion of equality for terms which we could prove equivalent to propositional
equality. Although interesting in its own right, this section will focus
on more advanced use cases.


\subsection{Simulation lemma}\label{section:simulation}

A constraint mentioning all three relation transformers appears naturally when
we want to say that a semantics can simulate another one. For instance, renaming
is simulated by substitution: we simply have to restrict ourselves to environments
mapping variables to terms which happen to be variables.
More generally, given a semantics \AB{𝓢ᴬ} with values \AB{𝓥ᴬ} and computations
\AB{𝓒ᴬ} and a semantics \AB{𝓢ᴮ} with values \AB{𝓥ᴮ} and computations \AB{𝓒ᴮ},
we want to establish the constraints under which these two semantics yield
related computations provided they were called with environments of related values.

These constraints are packaged in a record type called \AR{Simulation} and
parametrised over the semantics as well as the notion of relatedness used
for values (given by a relation \AB{𝓥ᴿ}) and computations
(given by a relation \AB{𝓒ᴿ}).

\begin{agdasnippet}
  \ExecuteMetaData[Generic/Simulation.tex]{recsim}
\end{agdasnippet}

The two first constraints are self-explanatory: the operations
\ARF{th\textasciicircum{}𝓥} and \ARF{var} defined by each semantics
should be compatible with the notions of relatedness used for values and computations.

\begin{agdasnippet}
\addtolength{\leftskip}{\parindent}
  \ExecuteMetaData[Generic/Simulation.tex]{thR}
  \ExecuteMetaData[Generic/Simulation.tex]{varR}
\end{agdasnippet}

The third constraint is similarly simple: the algebras (\ARF{alg}) should take
related recursively evaluated subterms of respective types
\AF{⟦} \AB{d} \AF{⟧} (\AF{Kripke} \AB{𝓥ᴬ} \AB{𝓒ᴬ}) and
\AF{⟦} \AB{d} \AF{⟧} (\AF{Kripke} \AB{𝓥ᴮ} \AB{𝓒ᴮ}) to related computations.
The difficuly is in defining an appropriate notion of relatedness \AF{bodyᴿ}
for these recursively evaluated subterms.

\begin{agdasnippet}
\addtolength{\leftskip}{\parindent}
  \ExecuteMetaData[Generic/Simulation.tex]{algR}
\end{agdasnippet}

We can combine \AF{⟦\_⟧ᴿ} and \AF{Kripkeᴿ} to express the idea that two recursively
evaluated subterms are related whenever they have an equal shape (which means their
Kripke functions can be grouped in pairs) and that all the pairs of Kripke function
spaces take related inputs to related outputs.

\begin{agdasnippet}
\addtolength{\leftskip}{\parindent}
  \ExecuteMetaData[Generic/Simulation.tex]{bodyR}
\end{agdasnippet}

The fundamental lemma of simulations is a generic theorem showing that for
each pair of \semrec{} respecting the \AR{Simulation} constraint, we
get related computations given environments of related input values.
This theorem is once more mutually
proven with a statement about \AF{Scope}s,
and \AD{Size}s play a crucial role in ensuring that the function is indeed total.

\begin{agdasnippet}
  \addtolength{\leftskip}{\parindent}
  \ExecuteMetaData[Generic/Simulation.tex]{simbody}
\end{agdasnippet}

Instantiating this generic simulation lemma, we can for instance prove
that renaming is a special case of substitution, or that renaming and
substitution are extensional, that is, that given environments equal in
a pointwise manner they produce syntactically equal terms. Of course these
results are not new but having them generically over all syntaxes with
binding is convenient. The first author experienced this first hand when tackling the
POPLMark Reloaded challenge~\citeyear{poplmarkreloaded} where
\AF{rensub} was actually needed.

\begin{agdasnippet}
  \ExecuteMetaData[Generic/Simulation/Syntactic.tex]{rensubfun}
  \ExecuteMetaData[Generic/Simulation/Syntactic.tex]{rensub}
\end{agdasnippet}

When studying specific languages, new opportunities to deploy the
fundamental lemma of simulations arise. The first author's solution
to the POPLMark Reloaded challenge
\citeyear{POPLMarkReloaded2019}
for instance describes the fact that
{(\AF{sub} \AB{$\rho$} \AB{t})}
reduces to {(\AF{sub} \AB{$\rho$'} \AB{t})} whenever for all \AB{v},
\AB{$\rho$}(\AB{v}) reduces to \AB{$\rho$'}(\AB{v}) as a \AR{Simulation}.
The main theorem (strong normalisation of STLC via a logical relation)
is itself an instance of (the unary version of) the simulation lemma.

The \AR{Simulation} proof framework is the simplest example of the abstract
proof frameworks introduced in ACMM~\citeyear{allais2017type}. We also
explain how a similar framework can be defined
for fusion lemmas and deploy it for the renaming-substitution interactions
but also their respective interactions with normalisation by evaluation.
Now that we are familiarised with the techniques at hand, we can tackle
this more complex example for all syntaxes definable in our framework.


\subsection{Fusion lemma}\label{section:fusion}

Results that can be reformulated as the ability to fuse two traversals
obtained as \semrec{} into one abound. When claiming that \AF{Tm} is
a Functor, we have to prove that two successive renamings can be fused into
a single renaming where the \AF{Thinning}s have been composed. Similarly,
demonstrating that \AF{Tm} is a relative Monad~\cite{JFR4389} implies proving
that two consecutive substitutions can be merged into a single one whose
environment is the first one, where the second one has been applied in a
pointwise manner. The \emph{Substitution Lemma} central
to most model constructions \cite{mitchell1991kripke} states
that a syntactic substitution followed by the evaluation of the resulting term
into the model is equivalent to the evaluation of the original term with an
environment corresponding to the evaluated substitution.

A direct application of these results is the first author's entry
\citeyear{POPLMarkReloaded2019}
to the
POPLMark Reloaded challenge~\citeyear{poplmarkreloaded}. Using a \AD{Desc}-based
representation of intrinsically well typed and well scoped terms we directly inherit
not only renaming and substitution but also all four fusion lemmas as corollaries
of our generic results. This allows us to remove the usual boilerplate
and go straight to the point.
As all of these statements have precisely the same structure, we can
once more devise a framework which will, provided that its constraints are
satisfied, prove a generic fusion lemma.

Fusion is more involved than simulation; we will once more step through
each one of the constraints individually, trying to give the reader an intuition
for why they are shaped the way they are.

\subsubsection{The fusion constraints}

The notion of fusion is defined for a triple of \AR{Semantics}; each \AB{𝓢ⁱ}
being defined for values in \AB{𝓥ⁱ} and computations in \AB{𝓒ⁱ}. The
fundamental lemma associated to such a set of constraints will state that
running \AB{𝓢ᴮ} after \AB{𝓢ᴬ} is equivalent to running \AB{𝓢ᴬᴮ} only.

The definition of fusion is parametrised by three relations: \AB{𝓔ᴿ} relates
triples of environments of values in {(\AB{Γ} \AR{─Env}) \AB{𝓥ᴬ} \AB{Δ}},
{(\AB{Δ} \AR{─Env}) \AB{𝓥ᴮ} \AB{Θ}} and {(\AB{Γ} \AR{─Env}) \AB{𝓥ᴬᴮ} \AB{Θ}}
respectively; \AB{𝓥ᴿ} relates pairs of values \AB{𝓥ᴮ} and \AB{𝓥ᴬᴮ};
and \AB{𝓒ᴿ}, our notion of equivalence for evaluation results, relates pairs
of computation in \AB{𝓒ᴮ} and \AB{𝓒ᴬᴮ}.

\begin{agdasnippet}
\ExecuteMetaData[Generic/Fusion.tex]{fusionrec}
\end{agdasnippet}
The first obstacle we face is the formal definition of \scarequote{running \AB{𝓢ᴮ}
after \AB{𝓢ᴬ}}: for this statement to make sense, the result of running
\AB{𝓢ᴬ} ought to be a term. Or rather, we ought to be able to extract a
term from a \AB{𝓒ᴬ}. Hence the first constraint: the existence of a \ARF{reifyᴬ}
function, which we supply as a field of the record \AR{Fusion}. When dealing with
syntactic semantics such as renaming or substitution
this function will be the identity. Nothing prevents proofs, such as the
idempotence of NbE, which use a bona fide reification function that extracts
terms from model values.

\begin{agdasnippet}
\addtolength{\leftskip}{\parindent}
\ExecuteMetaData[Generic/Fusion.tex]{reify}
\end{agdasnippet}
Then, we have to think about what happens when going under a binder: \AB{𝓢ᴬ}
will produce a \AF{Kripke} function space where a syntactic
value is required. Provided that \AB{𝓥ᴬ} is \AR{VarLike}, we can make use of \AF{reify}
to get a \AF{Scope} back. Hence the second constraint is:

\begin{agdasnippet}
\addtolength{\leftskip}{\parindent}
\ExecuteMetaData[Generic/Fusion.tex]{vlV}
\end{agdasnippet}
Still thinking about going under binders: if three evaluation environments
\AB{ρᴬ} in {(\AB{Γ} \AR{─Env}) \AB{𝓥ᴬ} \AB{Δ}}, \AB{ρᴮ} in
{(\AB{Δ} \AR{─Env}) \AB{𝓥ᴮ} \AB{Θ}}, and \AB{ρᴬᴮ} in {(\AB{Γ} \AR{─Env}) \AB{𝓥ᴬᴮ} \AB{Θ}}
are related by \AB{𝓔ᴿ} and we are given a thinning \AB{σ} from \AB{Θ} to \AB{Ω}
then \AB{ρᴬ}, the thinned \AB{ρᴮ} and the thinned \AB{ρᴬᴮ} should still be related.

\begin{agdasnippet}
\addtolength{\leftskip}{\parindent}
\ExecuteMetaData[Generic/Fusion.tex]{thV}
\end{agdasnippet}
Remembering that \AF{\_>>\_} is used in the definition of \AF{body} (Section~\ref{sec:fundamentallemma}) to
combine two disjoint environments {(\AB{Γ} \AR{─Env}) \AB{𝓥} \AB{Θ}} and
{(\AB{Δ} \AR{─Env}) \AB{𝓥} \AB{Θ}} into one of type
{((\AB{Γ} \AF{++} \AB{Δ}) \AR{─Env}) \AB{𝓥} \AB{Θ})}, we mechanically need a
constraint stating that \AF{\_>>\_} is compatible with \AB{𝓔ᴿ}. We demand
as an extra precondition that the values \AB{ρᴮ} and \AB{ρᴬᴮ} are extended
with are related according to \AB{𝓥ᴿ}. Lastly, for all the types to match up,
\AB{ρᴬ} has to be extended with placeholder variables which is possible because
we have already insisted on \AB{𝓥ᴬ} being \AR{VarLike}.

\begin{agdasnippet}
\addtolength{\leftskip}{\parindent}
\ExecuteMetaData[Generic/Fusion.tex]{appendR}
\end{agdasnippet}
We finally arrive at the constraints focusing on the semantical counterparts
of the terms' constructors. Each constraint essentially states that evaluating
a term with \AB{𝓢ᴬ}, reifying the result and running \AB{𝓢ᴮ} is equivalent to
using \AB{𝓢ᴬᴮ} straight away. This can be made formal by defining the following
relation \AF{𝓡}.

\begin{agdasnippet}
\addtolength{\leftskip}{\parindent}
\ExecuteMetaData[Generic/Fusion.tex]{crel}
\end{agdasnippet}
When evaluating a variable, on the one hand \AB{𝓢ᴬ}
will look up its meaning in the evaluation environment, turn the resulting value into
a computation which will get reified and then the result will be evaluated with \AB{𝓢ᴮ}.
Provided that all three evaluation environments are related by \AB{𝓔ᴿ} this should
be equivalent to looking up the value in \AB{𝓢ᴬᴮ}'s environment and turning it into a
computation. Hence the constraint \ARF{varᴿ}:

\begin{agdasnippet}
\addtolength{\leftskip}{\parindent}
\ExecuteMetaData[Generic/Fusion.tex]{varR}
\end{agdasnippet}
The case of the algebra follows a similar idea albeit being more complex:
a term gets evaluated using \AB{𝓢ᴬ} and to be able to run \AB{𝓢ᴮ}
afterwards we need to recover a piece of syntax. This is possible if the
\AF{Kripke} functional spaces are reified by being fed placeholder \AB{𝓥ᴬ} arguments
(which can be manufactured thanks to the \ARF{vl\^{𝓥ᴬ}} we mentioned before) and
then quoted. Provided that the result of running \AB{𝓢ᴮ} on that term is
related via \AF{⟦} \AB{d} \AF{⟧ᴿ} (\AF{Kripkeᴿ} \AB{𝓥ᴿ} \AB{𝓒ᴿ}) to the result
of running \AB{𝓢ᴬᴮ} on the original term, the \ARF{algᴿ} constraint states
that the two evaluations yield related computations.

\begin{agdasnippet}
\addtolength{\leftskip}{\parindent}
\ExecuteMetaData[Generic/Fusion.tex]{algR}
\end{agdasnippet}

\subsubsection{The fundamental lemma of fusion}

This set of constraints is enough to prove a fundamental lemma of \AR{Fusion}
stating that from a triple of related environments, one gets a pair of related
computations: the composition of \AB{𝓢ᴬ} and \AB{𝓢ᴮ} on one hand and
\AB{𝓢ᴬᴮ} on the other. This lemma is once again proven mutually with its
counterpart for \semrec{}'s \AF{body}'s action on \AR{Scope}s.

\begin{agdasnippet}
 \ExecuteMetaData[Generic/Fusion.tex]{fusiontype}
\end{agdasnippet}

\subsubsection{Instances of fusion}

A direct consequence of this result is the four lemmas collectively stating
that any pair of renamings and / or substitutions can be fused together to
produce either a renaming (in the renaming-renaming interaction case) or a
substitution (in all the other cases). One such example is the fusion of
substitution followed by renaming into a single substitution where the
renaming has been applied to the environment.

\begin{agdasnippet}
 \ExecuteMetaData[Generic/Fusion/Syntactic.tex]{subren}
\end{agdasnippet}

Another corollary of the fundamental lemma of fusion is the observation that
Kaiser, Schäfer, and Stark~\citeyear{Kaiser-wsdebr} make: \emph{assuming
functional extensionality}, all the ACMM~\citeyear{allais2017type} traversals
are compatible with variable renaming.
We reproduced this result generically for all syntaxes (see accompanying code).
The need for functional extensionality arises in the proof when dealing with
subterms which have extra bound variables. These terms are interpreted as
Kripke functional spaces in the host language and we can only prove that they
take equal inputs to equal outputs. An intensional notion of equality will
simply not do here.
As a consequence, we refrain from using the generic result in practice when
an axiom-free alternative is provable. Kaiser, Schäfer and Stark's observation
naturally raises the question of whether the same semantics are also stable
under substitution. Our semantics implementing printing with names is a clear
counterexample.


\subsection{Definition of bisimilarity for cofinite objects}

Although we were able to use propositional equality when studying
syntactic traversals working on terms, it is not the appropriate
notion of equality for cofinite trees. What we want is a generic
coinductive notion of bisimilarity for all cofinite tree types
obtained as the unfolding of a description. Two trees are bisimilar
if their top layers have the same shape and their substructures are
themselves bisimilar. This is precisely the type of relation \AF{⟦\_⟧ᴿ}
was defined to express. Hence the following coinductive relation.

\begin{agdasnippet}
 \ExecuteMetaData[Generic/Bisimilar.tex]{bisim}
\end{agdasnippet}

We can then prove by coinduction that this generic definition always gives
rise to an equivalence relation using the relator's stability properties
(if \AB{R} is reflexive / symmetric / transitive then so is {(\AF{⟦} \AB{d} \AF{⟧ᴿ} \AB{R})}
mentioned in Section~\ref{lem:zipstable}.

\begin{agdasnippet}
 \ExecuteMetaData[Generic/Bisimilar.tex]{eqrel}
\end{agdasnippet}

This definition can be readily deployed to prove, for example, that the unfolding
of \AF{01↺} defined in Section~\ref{def:colist} is indeed bisimilar to \AF{01⋯}
which was defined in direct style. The proof is straightforward due to the simplicity
of this example: the first \AIC{refl} witnesses the fact that both definitions
pick the same constructor (a cons cell), the second that they carry the
same natural number, and we can conclude by an appeal to the coinduction
hypothesis.

\begin{agdasnippet}
\ExecuteMetaData[Generic/Examples/Colist.tex]{bisim01}
\end{agdasnippet}



\section{Related work}\label{section:related-work}

\subsection{Variable binding} The representation of variable binding
in formal systems has been a hot topic for decades. Part of the purpose
of the first POPLMark challenge~\citeyear{poplmark} was to explore and
compare various methods.

Having based our work on a de Bruijn encoding of variables, and thus a
canonical treatment of \(\alpha\)-equivalence classes, our work has no
direct comparison with permutation-based treatments such as those of
Pitts' and Gabbay's nominal syntax~\citeyear{gabbay:newaas-jv}.

Our generic universe of syntax is based on
scoped and typed de Bruijn indices~\cite{de1972lambda} but it is not
a necessity. It is for instance possible to give an interpretation
of \AD{Desc}riptions corresponding to Chlipala's Parametric Higher-Order
Abstract Syntax~\citeyear{DBLP:conf/icfp/Chlipala08} and we would be interested
to see what the appropriate notion of \AD{Semantics} is for this representation.

\subsection{Alternative binding structures} The binding structure we
present here is based on a flat, lexical scoping strategy. There are
other strategies and it would be interesting to see whether
our approach could be reused in these cases.

Weirich, Yorgey, and Sheard's work~\citeyear{DBLP:conf/icfp/WeirichYS11}
encompassing a large array of patterns (nested, recursive, telescopic, and
n-ary) can inform our design. They do not enforce scoping invariants internally
which forces them to introduce separate constructors for a simple binder, a
recursive one, or a telescopic pattern. They recover guarantees by giving
their syntaxes a nominal semantics thus bolting down the precise meaning of
each combinator and then proving that users may only generate well formed
terms.

Bach Poulsen, Rouvoet, Tolmach, Krebbers and Visser~\citeyear{BachPoulsen}
introduce notions of scope graphs and frames to scale the techniques typical
of well scoped and typed deep embeddings to imperative languages.
They showcase the core ideas of their work using STLC extended with references
and then demonstrate that they can already handle a large subset of Middleweight
Java.
We have demonstrated that our framework could be used to define effectful
semantics by choosing an appropriate monad stack~\cite{DBLP:journals/iandc/Moggi91}.
This suggests we should be able to model STLC+Ref. It is however clear that
the scoping structures handled by scope graphs and frames are, in their full
generality, out of reach for our framework. In constrast, our work shines by
its generality: we define an entire universe of syntaxes and provide users
with traversals and lemmas implemented \emph{once and for all}.

Many other opportunities to enrich the notion of binder in our library are
highlighted by Cheney~\citeyear{DBLP:conf/icfp/Cheney05a}. As we have demonstrated
in Sections~\ref{section:letbinding} and \ref{section:inlining} we can already
handle let-bindings generically for all syntaxes. We are currently considering
the modification of our system to handle deeply nested patterns by removing the
constraint that the binders' and variables' sorts are identical. A notion of
binding corresponding to hierarchical namespaces would be an exciting addition.

We have demonstrated how to write generic programs over the potentially
cyclic structures of Ghani, Hamana, Uustalu and Vene~\citeyear{ghani2006representing}.
Further work by Hamana~\citeyear{Hamana2009} yielded a different presentation
of cyclic structures which preserves sharing: pointers can not only refer
to nodes above them but also across from them in the cyclic tree. Capturing
this class of inductive types as a set of syntaxes with binding and writing
generic programs over them is still an open problem.

\subsection{Semantics of syntaxes with binding} An early foundational study
of a general \emph{semantic} framework for signatures with binding, algebras
for such signatures, and initiality of the term algebra, giving rise to a
categorical \scarequote{program} for substitution and proofs of its properties, was given
by Fiore, Plotkin and Turi~\cite{FiorePlotkinTuri99}. They worked in the category of presheaves
over renamings, (a skeleton of) the category of finite sets. The presheaf
condition corresponds to our notion of being \AF{Thinnable}. Exhibiting
algebras based on both de Bruijn \emph{level} and \emph{index} encodings,
their approach isolates the usual (abstract) arithmetic required of such encodings.

By contrast, we are working in an \emph{implemented} type theory where the
encoding can be understood as its own foundation without appeal to an external
mathematical semantics. We are able to go further in developing machine-checked
such implementations and proofs, themselves generic with respect to an abstract syntax
\AD{Desc} of syntaxes with binding. Moreover, the usual source of implementation
anxiety, namely concrete arithmetic on de Bruijn indices, has been successfully
encapsulated via the \AF{□} coalgebra structure. It is perhaps noteworthy that
our type-theoretic constructions, by contrast with their categorical ones,
appear to make fewer commitments as to functoriality, thinnability, etc. in our
specification of semantics, with such properties typically being \emph{provable}
as a further instance of our framework.

\subsection{Meta-theory automation via tactics and code generation} The
tediousness of repeatedly
proving similar statements has unsurprisingly led to various attempts at
automating the pain away via either code generation or the definition of
tactics. These solutions can be seen as untrusted oracles driving the
interactive theorem prover.

Polonowski's DBGen~\citeyear{polonowski:db} takes as input a raw syntax with
comments annotating binding sites. It generates a module defining lifting,
substitution as well as a raw syntax using names and a validation function
transforming named terms into de Bruijn ones; we refrain from calling it a
scope checker as terms are not statically proven to be well scoped.

Kaiser, Schäfer, and Stark~\citeyear{Kaiser-wsdebr} build on our previous paper
to draft possible theoretical foundations for Autosubst, a so-far untrusted
set of tactics. The paper is based on a specific syntax: well scoped call-by-value
System F. In contrast, our effort has been here to carve out
a precise universe of syntaxes with binding and give a systematic account
of these syntaxes' semantics and proofs.

Keuchel, Weirich, and Schrijvers' Needle~\citeyear{needleandknot} is a code
generator written in Haskell producing syntax-specific Coq modules
implementing common traversals and lemmas about them.

\subsection{Universes of syntaxes with binding} Keeping in mind Altenkirch
and McBride's observation that generic programming is everyday programming
in dependently typed languages~\citeyear{DBLP:conf/ifip2-1/AltenkirchM02}, we can naturally
expect generic, provably sound, treatments of these notions in tools such as
Agda or Coq.

Keuchel~\citeyear{Keuchel:Thesis:2011} together with Jeuring~\citeyear{DBLP:conf/icfp/KeuchelJ12}
define a universe of syntaxes with binding with a rich notion of binding patterns
closed under products but also sums as long as the disjoint patterns bind the same
variables. They give their universe two distinct semantics: a first one based on well
scoped de Bruijn indices and a second one based on Parametric Higher-Order Abstract
Syntax (PHOAS)~\cite{DBLP:conf/icfp/Chlipala08} together with a generic conversion
function from the de Bruijn syntax to the PHOAS one. Following McBride's unpublished 2005 manuscript, which emerged as \cite{benton2012strongly},
they implement both renaming and substitution in one fell swoop. They leave other
opportunities for generic programming and proving to future work.

Keuchel, Weirich, and Schrijvers' Knot~\citeyear{needleandknot} implements
as a set of generic programs the traversals and lemmas generated in specialised
forms by their Needle program. They see Needle as a pragmatic choice: working
directly with the free monadic terms over finitary containers would be too cumbersome. In
the first author's experience solving the POPLMark Reloaded challenge, Agda's pattern
synonyms make working with an encoded definition almost
seamless.

The GMeta generic framework~\citeyear{gmeta} provides a universe of syntaxes
and offers various binding conventions (locally nameless~\cite{Chargueraud2012}
or de Bruijn indices). It also generically implements common traversals (e.g. computing
the sets of free variables,
shifting
de Bruijn indices or substituting terms for parameters) as well as common
predicates (e.g. being a closed term) and provides generic lemmas proving that
they are well behaved. It does not offer a generic framework
for defining new well scoped-and-typed semantics and proving their properties.

Érdi~\citeyear{gergodraft} defines a universe inspired by a first draft of this
paper and gives three different interpretations (raw, scoped and typed syntax)
related via erasure. He provides type- and scope-preserving renaming and
substitution as well as various generic proofs that they are well behaved but
offers neither a generic notion of semantics, nor generic proof frameworks.

Copello~\citeyear{copello2017} works with \emph{named} binders and
defines nominal techniques (e.g. name swapping) and ultimately $\alpha$-equivalence
over a universe of regular trees with binders inspired by Morris'~\citeyear{morris-regulartt}.

\subsection{Fusion of successive traversals}

The careful characterisation of the successive recursive traversals which can be
fused together into a single pass in a semantics-preserving way is not new. This
transformation is a much needed optimisation principle in a high-level functional
language.

Through the careful study of the recursion operator associated to each
strictly positive data type,
Malcolm~\citeyear{DBLP:journals/scp/Malcolm90} defined optimising
fusion proof principles.
Other optimisations such as deforestation~\cite{DBLP:journals/tcs/Wadler90}
or the compilation of a recursive definition into an equivalent abstract
machine-based tail-recursive program~\cite{DBLP:conf/icfp/CortinasS18}
rely on similar generic proofs that these transformations are meaning-preserving.


\section{Conclusion and future work}

Recalling our earlier work~\citeyear{allais2017type}
we have started from an example of a type- and scope-safe language (the simply typed
λ-calculus), have studied common invariant preserving traversals and noticed their
similarity. After introducing a notion of semantics and refactoring these traversals as
instances of the same fundamental lemma, we have observed the tight
connection between the abstract definition of semantics and the shape of the
language.

By extending a universe of data type descriptions to support a notion of binding,
we have given a generic presentation of syntaxes with binding. We then described
a large class of type- and scope-safe generic programs acting on all of them.
We started with syntactic traversals such as renaming and substitution. We then
demonstrated how to write a small compiler pipeline: scope checking, type checking
and elaboration to a core language, desugaring of new constructors added by a language
transformer, dead code elimination and inlining, partial evaluation, and printing
with names.

We have seen how to construct generic proofs about these generic programs. We
first introduced a Simulation relation showing what it means for two semantics
to yield related outputs whenever they are fed related input environments. We
then built on our experience to tackle a more involved case: identifying a set
of constraints guaranteeing that two semantics run consecutively can be subsumed
by a single pass of a third one.

We have put all of these results into practice using them to solve the
POPLMark Reloaded challenge
\citeyear{POPLMarkReloaded2019}
which consists of formalising strong
normalisation for the simply typed λ-calculus via a logical relation
argument. This also gave us the opportunity to try our framework on larger
languages by tackling the challenge's extensions to sum types and G\"{o}del's
System T.

Finally, we have demonstrated that this formalisation can be reused
in other domains by seeing our syntaxes with binding as potentially cyclic
terms. Their unfolding is a non-standard semantics and we provide the
user with a generic notion of bisimilarity to reason about them.

\subsection{Limitations of the current framework}

Although quite versatile already our current framework has some limitations
which suggest avenues for future work. We list these limitations from easiest
to hardest to resolve. Remember that each modification to the universe of
syntaxes needs to be given an appropriate semantics.

\paragraph*{Closure under products.} Our current universe of descriptions is
closed under sums as demonstrated in Section~\ref{desccomb}. It is however
not closed under products: two arbitrary right-nested products conforming
to a description may disagree on the sort of the term they are constructing.
An approach where the sort is an input from which the description of allowed
constructors is computed (à la Dagand \citeyear{DBLP:phd/ethos/Dagand13} where,
for instance, the \AIC{`lam} constructor is only offered if the input sort is
a function type) would not suffer from this limitation.

\paragraph*{Unrestricted variables.} Our current notion of variable can be used
to form a term of any sort. We remarked in Sections~\ref{section:typechecking}
and \ref{section:elaboration} that in some languages we want to restrict this
ability to one sort in particular. In that case, we wanted users to only be able
to use variables at the sort \AIC{Infer} of our bidirectional language. For the
time-being we made do by restricting the environment values our \AR{Semantics}
use to a subset of the sorts: terms with variables of the wrong sort will not be
given a semantics.

\paragraph*{Flat binding structure.} Our current set-up limits us to flat binding
structures: variables and binders share the same sorts. This prevents us from
representing languages with binding patterns, for instance pattern-matching
let-binders which can have arbitrarily nested patterns taking pairs apart.

\paragraph*{Closure under derivation.} One-hole contexts play a major role in the
theory of programming languages. Just like the one-hole context of a data type is
a data type~\cite{DBLP:journals/fuin/AbbottAMG05}, we would like our universe to
be closed under derivatives so that the formalisation of, for example, evaluation contexts
could benefit directly from the existing machinery.

\paragraph*{Closure under closures.} Jander's work on formalising and certifying
continuation-passing style transformations~\cite{Jander:Thesis:2019}
highlighted the need for a notion of syntaxes with closures. Recalling
that our notion of Semantics is always compatible with precomposition
with a renaming~\cite{Kaiser-wsdebr} but not necessarily
precomposition with a substitution (printing is, for instance, not
stable under substitution), accommodating terms with suspended
substitutions is a real challenge. Preliminary experiments show that a
drastic modification of the type of the fundamental lemma of
\AR{Semantics} makes dealing with such closures possible. Whether the
resulting traversal has good properties that can be proven generically
is still an open problem.

\subsection{Future work}

The diverse influences leading to this work suggest many opportunities for
future research.

\begin{itemize}
\item Our example of elaborating an enriched language to a core
  one, ACMM's implementation of a continuation-passing style
  conversion function, and Jander's work~\citeyear{Jander:Thesis:2019}
  on the certification of a intrinsically typed CPS transformation
  raises the question of how many such common compilation passes can
  be implemented generically.
\item Our universe only includes syntaxes that allow unrestricted
  variable use. Variables may be used multiple times or never, with no
  restriction. We are interested in representing syntaxes that only
  allow single use of variables, such as term calculi for linear logic
  \cite{DBLP:conf/tlca/BentonBPH93,barber96dual},
  or that annotate variables with usage information
  \cite{BrunelGMZ14,GhicaS14,PetricekOM14,context-constrained}, or arrange variables into
  non-list-like structures such as bunches
  \cite{DBLP:journals/jfp/OHearn03}, or arbitrary algebraic structures
  \cite{DBLP:conf/rta/LicataSR17}, and in investigating what form a
  generic semantics for these syntaxes takes.
\item An extension of Dagand and McBride's theory of
  ornaments~\citeyear{DBLP:journals/jfp/DagandM14} could provide an
  appropriate framework to formalise and mechanise the connection
  between various languages, some being seen as refinements of
  others. This is particularly evident when considering the
  informative type checker (see the accompanying code) which given a
  scoped term produces a scoped-and-typed term by type checking or
  type inference.
\item The first author's work on the POPLMark Reloaded challenge highlights a need
  for generic notions of congruence closure which would come with
  guarantees (if the original relation is stable under renaming and
  substitution so should the closure).  Similarly, the \scarequote{evaluation
  contexts} corresponding to a syntax could be derived automatically
  by building on the work of Huet~\citeyear{huet_1997} and Abbott,
  Altenkirch, McBride and
  Ghani~\citeyear{DBLP:journals/fuin/AbbottAMG05}, allowing us to
  revisit previous work based on concrete instances of ACMM such as
  McLaughlin, McKinna and Stark~\citeyear{craig2018triangle}.
\end{itemize}

We now know how to generically describe syntaxes and their well
behaved semantics. We can now start asking what it means to define
well behaved judgments. Why stop at helping the user write their
specific language's meta-theory when we could study meta-meta-theory?

%% file: catalogue/printing.tex
\subsection{Printing with names}\label{section:genericprinting}

We have seen in Section~\ref{section:printing} that printing with names
is an instance of ACMM's notion of \semrec{}. We will now show that this
observation can be generalised to arbitrary syntaxes with binding. Unlike
renaming or substitution, this generic program will require user guidance:
there is no way for us to guess how an encoded term should be printed. We
can however take care of the name generation (using the \AF{Fresh} monad from Page~\pageref{section:printing}), deal with variable binding,
and implement the traversal generically. We want our printer to have type:
\begin{agdasnippet}
\ExecuteMetaData[Generic/Semantics/Printing.tex]{printtype}
\end{agdasnippet}
where \AF{Display} explains how to print one `layer' of term provided that
we are handed the \AF{Pieces} corresponding to the printed subterm and
names for the bound variables:
\begin{agdasnippet}
  \ExecuteMetaData[Generic/Semantics/Printing.tex]{display}
\end{agdasnippet}
Reusing the notion of \AR{Name} introduced in Section~\ref{section:printing},
we can make \AF{Pieces} formal. A subterm has already been printed if we
have a string representation of it together with an environment of \AR{Name}s
we have attached to the newly bound variables this structure contains.
That is to say:
\begin{agdasnippet}
\ExecuteMetaData[Generic/Semantics/Printing.tex]{pieces}
\end{agdasnippet}
The key observation that will help us define a generic printer is that
\AF{Fresh} composed with \AR{Name} is \AR{VarLike}. Indeed, as the composition
of a functor and a trivially thinnable \AR{Wrap}per, \AF{Fresh} is \AF{Thinnable},
and \AF{fresh} (defined on Page~\pageref{section:printing}) is the proof that we
can generate placeholder values thanks to the name supply.

\begin{agdasnippet}
\ExecuteMetaData[Generic/Semantics/Printing.tex]{vlmname}
\end{agdasnippet}

This \AR{VarLike} instance empowers us to reify in an effectful manner
a \AF{Kripke} function space taking \AF{Name}s and returning a \AF{Printer}
to a set of \AF{Pieces}.

\begin{agdasnippet}
\ExecuteMetaData[Generic/Semantics/Printing.tex]{reifytype}
\end{agdasnippet}

In case there are no newly bound variables, the \AF{Kripke} function space
collapses to a mere \AR{Printer} which is precisely the wrapped version of
the type we expect.

\begin{agdasnippet}
\ExecuteMetaData[Generic/Semantics/Printing.tex]{reifybase}
\end{agdasnippet}

Otherwise we proceed in a manner reminiscent of the pure reification function
defined at the end of Section~\ref{section:renandsub}. We start by generating an environment
of names for the newly bound variables by using the fact that \AF{Fresh} composed
with \AF{Name} is \AR{VarLike} together with the fact that environments are
Traversable~\cite{mcbride_paterson_2008}, 
and thus admit the standard Haskell-like \AF{mapA} and \AF{sequenceA}
traversals. 
We then run the \AF{Kripke} function
on these names to obtain the string representation of the subterm. We finally
return the names we used together with this string.

\begin{agdasnippet}
  \ExecuteMetaData[Generic/Semantics/Printing.tex]{reifypieces}
\end{agdasnippet}

We can put all of these pieces together to obtain the \AF{Printing} semantics.
The first two constraints can be trivially discharged. When defining the
algebra we start by reifying the subterms, then use the fact that  one \scarequote{layer}
of term of our syntaxes with binding is always traversable to combine all of
these results into a value we can apply our display function to.

\begin{agdasnippet}
  \ExecuteMetaData[Generic/Semantics/Printing.tex]{printing}
\end{agdasnippet}

This allows us to write a \AF{printer} for open terms:

\begin{agdasnippet}
  \ExecuteMetaData[Generic/Semantics/Printing.tex]{print}
\end{agdasnippet}
We start by using \AF{base} (defined in Section~\ref{sec:varlike:base})
to generate an environment of \AR{Name}s for the free variables, then use
our semantics to get a \AF{printer} which we can run using a stream \AF{names} of distinct
strings as our name supply.

\paragraph*{Untyped λ-calculus.} Defining a printer for the untyped
λ-calculus is now very easy: we define a \AF{Display} by case analysis.
In the application case, we combine the string representation of the
function, wrap its argument's representation between parentheses and
concatenate the two together. In the lambda abstraction case, we are
handed the name the bound variable was assigned together with the body's
representation; it is once more a matter of putting the \AF{Pieces}
together.

\begin{agdasnippet}
\ExecuteMetaData[Generic/Examples/Printing.tex]{printUTLC}
\end{agdasnippet}

As always, these functions are readily executable and we can check
their behaviour by writing tests. First, we print the identity function
defined in Section~\ref{section:universe}
in an empty context and verify that we do obtain the string \AStr{"λa. a"}.
Next, we print an open term in a context of size two and can immediately
observe that names are generated for the free variables first, and then the
expression itself is printed.

  \begin{agdasnippet}
  \ExecuteMetaData[Generic/Examples/Printing.tex]{printid}
  \end{agdasnippet}
  \begin{agdasnippet}
  \ExecuteMetaData[Generic/Examples/Printing.tex]{printopen}
  \end{agdasnippet}

%% file: catalogue/scopechecking.tex
\subsection{Writing a generic scope checker}\label{section:genericscoping}

Converting terms in the internal syntax to strings which can in turn be
displayed in a terminal or an editor window is only part of a compiler's
interaction loop. The other direction takes strings as inputs and attempts to
produce terms in the internal syntax. The first step is to parse the input
strings into structured data, the second is to perform scope checking,
and the third step consists of type checking.

Parsing is currently out of scope for our library; users can write safe
ad-hoc parsers for their object language by either using a library of total
parser combinators~\cite{DBLP:conf/icfp/Danielsson10,allais2018agdarsec}
or invoking a parser generator oracle whose target is a total
language~\cite{Stump:2016:VFP:2841316}. As we will see shortly, we can
write a generic scope checker transforming terms in a raw syntax where
variables are represented as strings into a well scoped syntax. We will
come back to type checking with a concrete example in section~\ref{section:typechecking}
and then discuss related future work in the conclusion.

Our scope checker will be a function taking two explicit arguments: a
name for each variable in scope \AB{Γ} and a raw term for a syntax
description \AB{d}.  It will either fail (the Monad \AF{Fail} granting
us the ability to fail is defined below) or return a well scoped and
sorted term for that description.

\begin{agdasnippet}
\ExecuteMetaData[Generic/Scopecheck.tex]{totmtype}
\end{agdasnippet}

\paragraph*{Scope.} We can obtain \AF{Names}, the datastructure associating to
each variable in scope its raw name as a string by reusing the standard library's
\AD{All}. The inductive family \AD{All} is a predicate transformer making sure a
predicate holds of all the element of a list. It is defined in a style common in
Agda: because \AD{All}'s constructors are in one to one correspondence with that
of its index type (\AD{List} \AB{A}), the same name are reused: \AIC{[]} is the
name of the proof that \AB{P} trivially holds of all the elements in the empty
list \AIC{[]}; similarly \AIC{\_∷\_} is the proof that provided that \AB{P} holds
of the element \AB{a} on the one hand and of the elements of the list \AB{as}
on the other then it holds of all the elements of the list (\AB{a} \AIC{∷} \AB{as}).

\noindent
\begin{minipage}{\textwidth}
\begin{minipage}[t]{0.64\textwidth}
\ExecuteMetaData[Stdlib.tex]{all}
\end{minipage}
\begin{minipage}[t]{0.35\textwidth}
  \ExecuteMetaData[Generic/Scopecheck.tex]{names}
\end{minipage}
\end{minipage}

\paragraph*{Raw terms.}
The definition of \AF{WithNames} is analogous to \AF{Pieces} in the
previous section: we expect \AF{Names} for the newly bound
variables. Terms in the raw syntax then leverage these
definitions. They are either a variables or another \scarequote{layer} of raw
terms. Variables \AIC{'var} carry a \AD{String} and potentially some
extra information \AB{E} (typically a position in a file). The other
constructor \AIC{'con} carries a layer of raw terms where subterms are
raw terms equiped with names for any newly bound variables.

\begin{agdasnippet}
  \ExecuteMetaData[Generic/Scopecheck.tex]{withnames}
  \ExecuteMetaData[Generic/Scopecheck.tex]{raw}
\end{agdasnippet}

\paragraph*{Error handling.} Various things can go wrong during scope checking:
evidently a name can be out of scope but it is also possible that it may be
associated to a variable of the wrong sort. We define an enumerating type
covering these two cases. The scope checker will return a computation in the
Monad \AF{Fail} thus allowing us to fail and return an error, the string that
caused the failure and the extra data of type \AB{E} that accompanied it.

\noindent
\begin{minipage}{\textwidth}
\begin{minipage}[t]{0.5\textwidth}
  \ExecuteMetaData[Generic/Scopecheck.tex]{error}
\end{minipage}
\begin{minipage}[t]{0.4\textwidth}
  \ExecuteMetaData[Generic/Scopecheck.tex]{monad}
  \ExecuteMetaData[Generic/Scopecheck.tex]{fail}
\end{minipage}
\end{minipage}

Equipped with these notions, we can write down the type of \AF{toVar} which
tackles the core of the problem: variable resolution. The function takes a
string and a sort as well the names and sorts of the variables in the ambient
scope. Provided that we have a function \AB{\_≟I\_} to decide equality on sorts,
we can check whether the string corresponds to an existing variable and whether
that binding is of the right sort. Thus we either fail or return a well scoped
and well sorted \AD{Var}.

If the ambient scope is empty then we can only fail with an \AIC{OutOfScope} error.
Alternatively, if the variable's name corresponds to that of the first one
in scope we check that the sorts match up and either return \AIC{z} or fail
with a \AIC{WrongSort} error. Otherwise we look for the variable further
down the scope and use \AIC{s} to lift the result to the full scope.

\begin{agdasnippet}
\ExecuteMetaData[Generic/Scopecheck.tex]{toVar}
\end{agdasnippet}

Scope checking an entire term then amounts to lifting this action on
variables to an action on terms. The error Monad \AF{Fail} is by
definition an Applicative and by design our terms are
Traversable~\cite{bird_paterson_1999,DBLP:journals/jfp/GibbonsO09}.
The action on term is defined mutually with the action on scopes.
As we can see in the second equation for \AF{toScope}, thanks to the
definition of \AF{WithNames}, concrete names arrive just in time to
check the subterm with newly bound variables.

\begin{agdasnippet}
  \ExecuteMetaData[Generic/Scopecheck.tex]{scopecheck}
\end{agdasnippet}

%% file: catalogue/typechecking.tex
\subsection{An algebraic approach to type checking}\label{section:typechecking}

Following Atkey~\citeyear{atkey2015algebraic}, we can consider type checking
and type inference as a possible semantics for a bidirectional~\cite{pierce2000local}
language. We reuse the syntax introduced in Section~\ref{par:bidirectional}
and the types introduced for the STLC at the end of Section~\ref{section:primer-term}; it
gives us a simply typed bidirectional calculus as a bisorted language using
a notion of \AD{Mode} to distinguish between terms for which we will be able to
\AIC{Infer} the type and the ones for which we will have to \AIC{Check} a type
candidate.

The values stored in the environment of the type checking function attach \AD{Type}
information to bound variables whose \AD{Mode} is \AIC{Infer}, guaranteeing no
variable ever uses the \AIC{Check} mode. In contrast, the generated computations
will, depending on the mode, either take a type candidate and \AIC{Check} it is
valid or \AIC{Infer} a type for their argument. These computations are always
potentially failing so we use the \AD{Maybe} monad.
In an actual compiler pipeline we would naturally use a different error monad
and generate helpful error messages pointing out where the type error occured. The
interested reader can see a fine-grained analysis of type errors in the extended
example of a type checker in \citet{DBLP:journals/jfp/McBrideM04}.

\noindent
\begin{minipage}{\textwidth}
\begin{minipage}[t]{0.40\textwidth}
  \ExecuteMetaData[Generic/Semantics/TypeChecking.tex]{varmode}
\end{minipage}\hfill
\begin{minipage}[t]{0.50\textwidth}
  \ExecuteMetaData[Generic/Semantics/TypeChecking.tex]{typemode}
\end{minipage}
\end{minipage}

A change of direction from inferring to checking will require being able to check
that two types agree so we introduce the function \AF{\_=?\_}. Similarly we will
sometimes expect a function type but may be handed anything so we will have to check
with \AF{isArrow} that our candidate's head constructor is indeed an arrow, and
collect the domain and codomain.

\noindent
\begin{minipage}{\textwidth}
\begin{minipage}[t]{0.53\textwidth}
  \ExecuteMetaData[Generic/Semantics/TypeChecking.tex]{typeeq}
\end{minipage}
\begin{minipage}[t]{0.46\textwidth}
  \ExecuteMetaData[Generic/Semantics/TypeChecking.tex]{isArrow}
\end{minipage}
\end{minipage}

We can now define type checking as a \semrec{}. We describe the algorithm constructor
by constructor; in the \AR{Semantics} definition (omitted here) the algebra will
simply perform a dispatch and pick the relevant auxiliary lemma. Note that in the
following code, \AF{\_<\$\_} is, following classic Haskell notations, the function
which takes an \AB{A} and a {\AD{Maybe} \AB{B}} and returns a {\AD{Maybe} \AB{A}}
which has the same structure as its second argument.

\paragraph*{Application.} When facing an application: infer the type of the function,
make sure it is an arrow type, check the argument at the domain's type and return
the codomain.
\begin{agdasnippet}
\ExecuteMetaData[Generic/Semantics/TypeChecking.tex]{app}
\end{agdasnippet}
\paragraph*{λ-abstraction.} For a λ-abstraction: check that the input
type \AB{arr} is an arrow type and check the body \AB{b} at the
codomain type in the extended environment (using \AF{bind}) where the
newly bound variable is of mode \AIC{Infer} and has the domain's type.
\begin{agdasnippet}
\ExecuteMetaData[Generic/Semantics/TypeChecking.tex]{lam}
\end{agdasnippet}
\paragraph*{Embedding of \AD{Infer} into \AD{Check}.} The change of
direction from \AIC{Infer}rable to \AIC{Check}able is successful when the
inferred type is equal to the expected one.
\begin{agdasnippet}
\ExecuteMetaData[Generic/Semantics/TypeChecking.tex]{emb}
\end{agdasnippet}
\paragraph*{Cut: A \AD{Check} in an \AD{Infer} position.}
So far, our bidirectional syntax only permits the construction
  of STLC terms in \emph{canonical
    form}~\cite{Pfenning:04,Dunfield:2004:TT:964001.964025}. In order to construct
  non-normal (redex) terms, whose semantics is given logically by the
  `cut' rule, we need to reverse direction.
Our final semantic operation, \AF{cut},
always comes with a type candidate against which to check the term and
to be returned in case of success.
\begin{agdasnippet}
\ExecuteMetaData[Generic/Semantics/TypeChecking.tex]{cut}
\end{agdasnippet}
We have defined a bidirectional type checker for this simple language by
leveraging the \semrec{} framework. We can readily run it on closed terms
using the \AF{closed} corollary defined in Section~\ref{sec:fundamentallemma}
and (defining \AF{β} to be {(\AIC{α} \AIC{`→} \AIC{α})}) infer the type of
the expression {(λx. x : β → β) (λx. x)}.

\begin{agdasnippet}
  \ExecuteMetaData[Generic/Semantics/TypeChecking.tex]{type-}
  \ExecuteMetaData[Generic/Semantics/TypeChecking.tex]{example}
\end{agdasnippet}

The output of this function is not very informative. As we will see shortly,
there is nothing stopping us from moving away from a simple computation
returning a {(\AD{Maybe} \AD{Type})} to an evidence-producing function
elaborating a term in \AF{Bidi} to a well scoped and typed term in \AF{STLC}.

%% file: catalogue/elaborating.tex
\subsection{An algebraic approach to elaboration}\label{section:elaboration}

Instead of generating a type or checking that a candidate will do, we
can use our language of \AD{Desc}riptions to define not only an
untyped source language but also an intrinsically typed internal
language. During type checking we simultaneously generate an
expression's type and a well scoped and well typed term of that
type. We use \AF{STLC} (defined in Section~\ref{par:intrinsicSTLC}) as
our internal language.

Before we can jump right in, we need to set the stage: a \AR{Semantics} for a
\AF{Bidi} term will involve ({\AD{Mode} \AF{─Scoped}}) notions of values and
computations but an \AF{STLC} term is ({\AD{Type} \AF{─Scoped}}). We first
introduce a \AF{Typing} associating types to each of the modes in scope,
together with an erasure function \AF{⌞\_⌟} extracting the context of types
implicitly defined by such a \AF{Typing}.
We will systematically distinguish contexts of modes (typically named \AB{ms})
and their associated typings (typically named \AB{Γ}).

\noindent
\begin{minipage}{\textwidth}
\begin{minipage}[t]{0.4\textwidth}
  \ExecuteMetaData[Generic/Semantics/Elaboration/Typed.tex]{typing}
\end{minipage}
\begin{minipage}[t]{0.5\textwidth}
  \ExecuteMetaData[Generic/Semantics/Elaboration/Typed.tex]{fromtyping}
\end{minipage}
\end{minipage}

We can then explain what it means for an elaboration process of type \AB{σ}
in a context of modes \AB{ms} to produce a term of the
({\AD{Type} \AF{─Scoped}}) family \AB{T}: for any typing \AB{Γ} of this
context of modes, we should get a value of type
{(\AB{T} \AB{σ} \AF{⌞} \AB{Γ} \AF{⌟})}.

\begin{agdasnippet}
  \ExecuteMetaData[Generic/Semantics/Elaboration/Typed.tex]{elab}
\end{agdasnippet}

Our first example of an elaboration process is our notion of environment values.
To each variable in scope of mode \AIC{Infer} we associate an elaboration function
targeting \AD{Var}. In other words: our values are all in scope i.e. provided any
typing of the scope of modes, we can assuredly return a type together with a
variable of that type.

\begin{agdasnippet}
  \ExecuteMetaData[Generic/Semantics/Elaboration/Typed.tex]{varmode}
\end{agdasnippet}

We can for instance prove that we have such an inference function for a newly bound
variable of mode \AIC{Infer}: given that the context has been extended with a variable
of mode \AIC{Infer}, the \AF{Typing} must also have been extended with a type \AB{σ}.
We can return that type paired with the variable \AIC{z}.

\begin{agdasnippet}
  \ExecuteMetaData[Generic/Semantics/Elaboration/Typed.tex]{var0}
\end{agdasnippet}

The computations are a bit more tricky. On the one hand, if we are in checking mode
then we expect that for any typing of the scope of modes and any type candidate we
can \AD{Maybe} return a term at that type in the induced context. On the other hand,
in the inference mode we expect that given any typing of the scope, we can \AD{Maybe}
return a type together with a term at that type in the induced context.

\begin{agdasnippet}
  \ExecuteMetaData[Generic/Semantics/Elaboration/Typed.tex]{elabmode}
\end{agdasnippet}

Because we are now writing a type checker which returns evidence of its claims, we need
more informative variants of the equality and \AF{isArrow} checks. In the equality
checking case we want to get a proof of propositional equality but we only care
about the successful path and will happily return \AIC{nothing} when failing.
Agda's support for (dependent!) \AK{do}-notation makes writing the check
really easy. For the arrow type, we introduce a family \AD{Arrow} constraining the
shape of its index to be an arrow type and redefine \AF{isArrow} as a \emph{view} targeting
this inductive family~\cite{DBLP:conf/popl/Wadler87,DBLP:journals/jfp/McBrideM04}.
We deliberately overload the constructor of the \AD{isArrow} family by calling
it \AIC{\_`→\_}. This means that the proof that a given type has the shape
{(\AB{σ} \AIC{`→} \AB{τ})} is literally written {(\AB{σ} \AIC{`→} \AB{τ})}.
This allows us to specify \emph{in the type} whether we want to work with the
full set of values in \AD{Type} or only the subset corresponding to function
types and to then proceed to write the same programs a Haskell programmers would,
with the added confidence that ours are guaranteed to be total.

\noindent
\begin{minipage}{\textwidth}
\begin{minipage}[t]{0.5\textwidth}
  \ExecuteMetaData[Generic/Semantics/Elaboration/Typed.tex]{equal}
\end{minipage}
\begin{minipage}[t]{0.45\textwidth}
  \ExecuteMetaData[Generic/Semantics/Elaboration/Typed.tex]{arrow}
\end{minipage}
\end{minipage}

We now have all the basic pieces and can start writing elaboration code. We
will use lowercase letter for terms in \AF{Bidi} and uppercase ones for their
elaborated counterparts in \AF{STLC}. We once more start by dealing with each
constructor in isolation before putting everything together to get a
\AR{Semantics}. These steps are very similar to the ones in the previous
section.

\paragraph*{Application.} In the application case, we start by elaborating the
function and we get its type together with its internal representation. We then
check that the inferred type is indeed an \AD{Arrow} and elaborate the argument
using the corresponding domain. We conclude by returning the codomain together
with the internal function applied to the internal argument.
\begin{agdasnippet}
  \ExecuteMetaData[Generic/Semantics/Elaboration/Typed.tex]{app}
\end{agdasnippet}
\paragraph*{λ-abstraction.} For the λ-abstraction case, we start by
checking that the type candidate \AB{arr} is an \AD{Arrow}. We can
then elaborate the body \AB{b} of the lambda in a context of modes extended
with one \AIC{Infer} variable, and the corresponding \AF{Typing} extended
with the function's domain. From this we get
an internal term \AB{B} corresponding to the body of the λ-abstraction and
conclude by returning it wrapped in a \AIC{`lam} constructor.
\begin{agdasnippet}
  \ExecuteMetaData[Generic/Semantics/Elaboration/Typed.tex]{lam}
\end{agdasnippet}
\paragraph*{Cut: A \AD{Check} in an \AD{Infer} position.} For cut, we start by
elaborating the term with the type annotation provided and return them paired
together.
\begin{agdasnippet}
  \ExecuteMetaData[Generic/Semantics/Elaboration/Typed.tex]{cut}
\end{agdasnippet}
\paragraph*{Embedding of \AD{Infer} into \AD{Check}.} For the change of direction
\AIC{Emb} we not only want to check that the inferred type and the type candidate
are equal: we need to cast the internal term labelled with the inferred type to
match the type candidate. Luckily, Agda's dependent \AK{do}-notation make our
job easy once again: when we make the pattern \AIC{refl} explicit, the equality holds
in the rest of the block.
\begin{agdasnippet}
  \ExecuteMetaData[Generic/Semantics/Elaboration/Typed.tex]{emb}
\end{agdasnippet}

We have almost everything we need to define elaboration as a semantics. Discharging
the \ARF{th\textasciicircum{}𝓥} constraint is a bit laborious and the proof doesn't
yield any additional insight so we leave it out here. The semantical counterpart of
variables (\ARF{var}) is fairly straightforward: provided a \AF{Typing}, we run the
inference and touch it up to return a term rather than a mere variable. Finally we
define the algebra (\ARF{alg}) by pattern-matching on the constructor and using our
previous combinators.

\begin{agdasnippet}
  \ExecuteMetaData[Generic/Semantics/Elaboration/Typed.tex]{elaborate}
\end{agdasnippet}

We can once more define a specialised version of the traversal induced by this
\AR{Semantics} for closed terms: not only can we give a (trivial) initial
environment (using the \AF{closed} corollary defined in Section~\ref{sec:fundamentallemma})
but we can also give a (trivial) initial \AF{Typing}. This leads to these
definitions:

\noindent
\begin{minipage}{\textwidth}
\begin{minipage}{0.55\textwidth}
  \ExecuteMetaData[Generic/Semantics/Elaboration/Typed.tex]{typemode}
\end{minipage}
\begin{minipage}{0.44\textwidth}
  \ExecuteMetaData[Generic/Semantics/Elaboration/Typed.tex]{type-}
\end{minipage}
\end{minipage}

Revisiting the example introduced in Section~\ref{section:typechecking},
we can check that elaborating the expression {(λx. x : β → β) (λx. x)}
yields the type {β} together with the term {(λx. x) (λx. x)} in internal
syntax. Type annotations have disappeared in the internal syntax as all
the type invariants are enforced intrinsically.

\begin{agdasnippet}
  \ExecuteMetaData[Generic/Semantics/Elaboration/Typed.tex]{example}
\end{agdasnippet}

%% file: catalogue/desugaring.tex
\subsection{Sugar and desugaring as a semantics}\label{section:letbinding}

One of the advantages of having a universe of programming language
descriptions is the ability to concisely define an \emph{extension}
of an existing language by using \AD{Desc}ription transformers
grafting extra constructors à la Swiestra~\citeyear{swierstra_2008}.
This is made extremely simple by the disjoint sum combinator
\AF{\_`+\_} which we defined in Section~\ref{section:universe}.
An example of such an extension is the addition of let-bindings to
an existing language.

let-bindings allow the user to avoid repeating themselves by naming
sub-expressions and then using these names to refer to the associated
terms. Preprocessors adding these types of mechanisms to existing
languages (from C to CSS) are rather popular. We introduce a
description \AD{Let} which can be used to extend any language
description \AB{d} to a language with let-bindings (\AB{d} \AF{`+}
\AF{Let}).

\noindent
\begin{minipage}{\textwidth}
\begin{minipage}[t]{0.45\textwidth}
  \ExecuteMetaData[Generic/Syntax/LetBinder.tex]{letcode}
\end{minipage}
\begin{minipage}[t]{0.45\textwidth}
  \ExecuteMetaData[Generic/Syntax/LetBinder.tex]{letpattern}
\end{minipage}
\end{minipage}

This description states that a let-binding node stores a pair of types
\AB{$\sigma$} and \AB{$\tau$} and two subterms. First comes the let-bound
expression of type \AB{$\sigma$} and second comes the body of the let which
has type \AB{$\tau$} in a context extended with a fresh variable of type
\AB{$\sigma$}. This defines a term of type \AB{$\tau$}.

In a dependently typed language, a type may depend on a value which
in the presence of let-bindings may be a variable standing for an
expression. The user naturally does not want it to make any difference
whether they used a variable referring to a let-bound expression or
the expression itself. Various type checking strategies can accommodate
this expectation: in Coq~\cite{Coq:manual} let-bindings are primitive
constructs of the language and have their own typing and reduction
rules whereas in Agda they are elaborated away to the core language
by inlining.

This latter approach to extending a language \AB{d} with let-bindings
by inlining them before type checking can be implemented generically as
a semantics over (\AB{d} \AF{`+} \AF{Let}). For this semantics values
in the environment and computations are both let-free terms. The algebra
of the semantics can be defined by parts thanks to \AF{case}, the eliminator
for \AF{\_`+\_} defined in Section~\ref{section:universe}:
the old constructors are kept the same by
interpreting them using the generic substitution algebra (\AF{Sub});
whilst the let-binder precisely provides the extra value to be added to the
environment.

\begin{agdasnippet}
  \ExecuteMetaData[Generic/Semantics/Elaboration/LetBinder.tex]{unletcode}
\end{agdasnippet}

The process of removing let-binders is then kickstarted with the placeholder
environment \AF{id\textasciicircum{}Tm}~=~\AIC{pack}~\AIC{`var} 
of type {(\AB{Γ} \AR{─Env}) (\AD{Tm} \AB{d} ∞) \AB{Γ}}. 

\begin{agdasnippet}
  \ExecuteMetaData[Generic/Semantics/Elaboration/LetBinder.tex]{unlet}
\end{agdasnippet}

In less than 10 lines of code we have defined a generic extension of
syntaxes with binding together with a semantics which corresponds
to an elaborator translating away this new construct.
In ACMM~\citeyear{allais2017type}, we focused on STLC only
and showed that it is similarly possible to implement a Continuation
Passing Style transformation as the composition of two semantics
à la Hatcliff and Danvy~\citeyear{hatcliff1994generic}.
The first semantics embeds STLC into Moggi's
Meta-Language~\citeyear{DBLP:journals/iandc/Moggi91} and thus fixes
an evaluation order. The second one translates Moggi's ML back into
STLC in terms of explicit continuations with a fixed return type.

We have demonstrated how easily one can define extensions and combine
them on top of a base language without having to reimplement common
traversals for each one of the intermediate representations. Moreover,
it is possible to define \emph{generic} transformations elaborating
these added features in terms of lower-level ones. This suggests that
this setup could be a good candidate to implement generic compilation
passes and could deal with a framework using a wealth of slightly
different intermediate languages à la Nanopass~\cite{Keep:2013:NFC:2544174.2500618}.

%% file: catalogue/inlining.tex
\subsection{Reference counting and inlining as a semantics}\label{section:inlining}

Although useful in its own right, desugaring all let-bindings can lead
to an exponential blow-up in code size. Compiler passes typically try
to maintain sharing by only inlining let-bound expressions which appear
at most one time. Unused expressions are eliminated as dead code whilst
expressions used exactly one time can be inlined: this transformation is
size preserving and opens up opportunities for additional optimisations.

As we will see shortly, we can implement reference counting and size
respecting let-inlining as a generic transformation over all syntaxes
with binding equipped with let-binders. This two-pass simple transformation
takes linear time which may seem surprising given the results due to Appel and
Jim~\citeyear{DBLP:journals/jfp/AppelJ97}. Our optimisation only inlines
let-bound variables whereas theirs also encompasses the reduction of static
β-redexes of (potentially) recursive function. While we can easily count how
often a variable is used in the body of a let-binder, the interaction between
inlining and β-reduction in theirs creates cascading simplification opportunities
thus making the problem much harder.

But first, we need to look at an example demonstrating that this is a
slightly subtle matter. Assuming that \AB{expensive} takes a long time
to evaluate, inlining all of the lets in the first expression is a really
good idea whilst we only want to inline the one binding \AB{y} in the
second one to avoid duplicating work. That is to say that the contribution
of the expression bound to \AB{y} in the overall count depends directly
on whether \AB{y} itself appears free in the body of the let which binds it.

\noindent
\begin{minipage}{\textwidth}
  \begin{minipage}{0.45\textwidth}
    \centering
    \ExecuteMetaData[Generic/Syntax/LetCounter.tex]{cheap}
  \end{minipage}
  \begin{minipage}{0.45\textwidth}
    \centering
    \ExecuteMetaData[Generic/Syntax/LetCounter.tex]{expensive}
  \end{minipage}
\end{minipage}

Our transformation will consist of two passes: the first one will annotate
the tree with accurate count information precisely recording whether
let-bound variables are used \AIC{zero}, \AIC{one}, or \AIC{many} times.
The second one will inline precisely the let-binders whose variable is
used at most once.

During the counting phase we need to be particularly careful not to overestimate
the contribution of a let-bound expression. If the let-bound variable is not used
then we can naturally safely ignore the associated count. But if it used \AIC{many}
times then we know we will not inline this let-binding and the count should
therefore only contribute once to the running total. We define the \AF{control}
combinator below precisely to explicitly handle this
subtle case.

The first step is to introduce the \AD{Counter} additive monoid. Addition will
allow us to combine counts coming from different subterms: if any of the two
counters is \AIC{zero} then we return the other, otherwise we know we have
\AIC{many} occurences.

\noindent
\begin{minipage}{\textwidth}
  \begin{minipage}{0.45\textwidth}
    \ExecuteMetaData[Generic/Syntax/LetCounter.tex]{counter}
  \end{minipage}
  \begin{minipage}{0.45\textwidth}
    \ExecuteMetaData[Generic/Syntax/LetCounter.tex]{addition}
  \end{minipage}
\end{minipage}

The syntax extension \AF{CLet} defined as follows is
a variation on the \AF{Let} syntax extension of Section~\ref{section:letbinding},
attaching a \AD{Counter} to each \AF{Let} node. The annotation process
can then be described as a function computing a
{(\AB{d} \AF{`+} \AF{CLet})} term from a {(\AB{d} \AF{`+} \AF{Let})} one.

\begin{agdasnippet}
  \ExecuteMetaData[Generic/Syntax/LetCounter.tex]{clet}
\end{agdasnippet}

We keep a tally of the usage information for the variables in scope. This
allows us to know which \AD{Counter} to attach to each \AF{Let} node.
Following the same strategy as in Section~\ref{section:genericscoping},
we use the standard library's \AD{All} to represent this mapping. We say
that a scoped value has been \AF{Counted} if it is paired with a \AD{Count}.

\noindent
\begin{minipage}{\textwidth}
  \begin{minipage}{0.45\textwidth}
    \ExecuteMetaData[Generic/Syntax/LetCounter.tex]{count}
  \end{minipage}
  \begin{minipage}{0.45\textwidth}
    \ExecuteMetaData[Generic/Semantics/Elaboration/LetCounter.tex]{counted}
  \end{minipage}
\end{minipage}

The two most basic counts are \AF{zeros} and \AF{fromVar}: the
empty one is \AIC{zero} everywhere and the one corresponding to a single use
of a single variable \AB{v} which is \AIC{zero} everywhere except for \AB{v}
where it is \AIC{one}.

\noindent
\begin{minipage}{\textwidth}
  \begin{minipage}{0.45\textwidth}
    \ExecuteMetaData[Generic/Syntax/LetCounter.tex]{zeros}
  \end{minipage}
  \begin{minipage}{0.45\textwidth}
    \ExecuteMetaData[Generic/Syntax/LetCounter.tex]{fromVar}
  \end{minipage}
\end{minipage}

When we collect usage information from different subterms, we need to put the
various counts together. The combinators we now define
allow us to easily do so: \AF{merge} adds up two counts in a pointwise manner
while \AF{control} uses one \AD{Counter} to decide whether to erase an existing
\AD{Count}. This is particularly convenient when computing the contribution of
a let-bound expression to the total tally: the contribution of the let-bound
expression will only matter if the corresponding variable is actually used.

\noindent
\begin{minipage}{\textwidth}
  \begin{minipage}{0.5\textwidth}
    \ExecuteMetaData[Generic/Syntax/LetCounter.tex]{merge}
  \end{minipage}
  \begin{minipage}{0.49\textwidth}
    \ExecuteMetaData[Generic/Syntax/LetCounter.tex]{control}
  \end{minipage}
\end{minipage}

We can now focus on the core of the annotation phase, defining a
\AR{Semantics} whose values are variables themselves and whose computations
are the pairing of a term in {(\AB{d} \AF{`+} \AF{CLet})} together with
a \AF{Count}. The variable case is trivial: provided a variable \AB{v},
we return {(\AIC{`var} \AB{v})} together with the count {(\AF{fromVar} \AB{v})}.

The non-let case is purely structural: we reify the \AF{Kripke} function
space and obtain a scope together with the corresponding \AF{Count}. We
unceremoniously \AF{drop} the \AD{Counter}s associated to the variables
bound in this subterm and return the scope together with the tally for
the ambient context.

\begin{agdasnippet}
  \ExecuteMetaData[Generic/Semantics/Elaboration/LetCounter.tex]{reifycount}
\end{agdasnippet}

The \AF{Let}-to-\AF{CLet} case is the most
interesting one. We start by reifying the \AB{body} of the let-binder which
gives us a tally \AB{cx} for the bound variable and \AB{ct} for the body's
contribution to the ambient environment's \AD{Count}. We annotate the node
with \AB{cx} and use it as a \AF{control} to decide whether we are going to
merge any of the let-bound's expression contribution \AB{ce} to form the
overall tally.

\begin{agdasnippet}
  \ExecuteMetaData[Generic/Semantics/Elaboration/LetCounter.tex]{letcount}
\end{agdasnippet}

Putting all of these things together we obtain the \AR{Semantics} \AF{Annotate}.
We promptly specialise it using an environment of placeholder values to obtain
the traversal \AF{annotate} elaborating raw let-binders into counted ones.

\begin{agdasnippet}
  \ExecuteMetaData[Generic/Semantics/Elaboration/LetCounter.tex]{annotate}
\end{agdasnippet}

Using techniques similar to the ones described in Section~\ref{section:letbinding},
we can write an \AF{Inline} semantics working on {(\AB{d} \AF{`+} \AF{CLet})} terms
and producing {(\AB{d} \AF{`+} \AF{Let})} ones. We make sure to preserve all the
let-binders annotated with \AIC{many} and to inline all the other ones. By composing
\AF{Annotate} with \AF{Inline} we obtain a size-preserving generic optimisation pass.

%% file: catalogue/normalising.tex
\subsection{(Unsafe) Normalisation by evaluation}\label{section:nbyeval}

A key type of traversal we have not studied yet is a language's
evaluator. Our universe of syntaxes with binding does not impose
any typing discipline on the user-defined languages and as such
cannot guarantee their totality. This is embodied by one of our running
examples: the untyped λ-calculus. As a consequence there
is no hope for a safe generic framework to define normalisation
functions.

The clear connection between the \AF{Kripke} functional space
characteristic of our semantics and the one that shows up in
normalisation by evaluation suggests we ought to manage to
give an unsafe generic framework for normalisation by evaluation.
By temporarily disabling Agda's positivity checker,
we can define a generic reflexive domain \AD{Dm} in which to
interpret our syntaxes. It has three constructors corresponding
respectively to a free variable, a constructor's counterpart where
scopes have become \AF{Kripke} functional spaces on \AD{Dm} and
an error token because the evaluation of untyped programs may go wrong.

\begin{agdasnippet}
  \ExecuteMetaData[Generic/Semantics/NbyE.tex]{domain}
\end{agdasnippet}

This data type definition is utterly unsafe. The more conservative
user will happily restrict themselves to particular syntaxes where
the typed settings allows for a domain to be defined as a logical
predicate or opt instead for a step-indexed approach.

But this domain does make it possible to define a generic \AF{nbe}
semantics which, given a term, produces a value in the reflexive
domain. Thanks to the fact we have picked a universe of finitary syntaxes, we
can \emph{traverse}~\cite{mcbride_paterson_2008,DBLP:journals/jfp/GibbonsO09}
the functor to define
a (potentially failing) reification function turning elements of the
reflexive domain into terms. By composing them, we obtain the
normalisation function which gives its name to normalisation by
evaluation.

The user still has to explicitly pass an interpretation of
the various constructors because there is no way for us to
know what the binders are supposed to represent: they may
stand for λ-abstractions, $\Sigma$-types, fixpoints, or
anything else.

\begin{agdasnippet}
  \ExecuteMetaData[Generic/Semantics/NbyE.tex]{nbe-setup}
\end{agdasnippet}

Using this setup, we can write a normaliser for the untyped λ-calculus
by providing an algebra. The key observation that allows us to implement
this algebra is that we can turn a Kripke function, \AB{f}, mapping values
of type \AB{σ} to computations of type \AB{τ} into an Agda function from
values of type \AB{σ} to computations of type \AB{τ}. This is witnessed
by the application function (\AF{\_\$\$\_}):
we first use \AF{extract}, defined in Section~\ref{sec:genenvironment}, to obtain
a function taking environments of values to computations. We then use the environment building
combinators defined there to manufacture the singleton
environment {(\AF{ε} \AB{∙} \AB{t})} containing the value \AB{t} of type
\AB{σ}.

\begin{agdasnippet}
  \ExecuteMetaData[Generic/Examples/NbyE.tex]{app}
\end{agdasnippet}

We now define two patterns for semantical values: one for application and
the other for lambda abstraction. This should make the case of interest of
our algebra (a function applied to an argument) fairly readable.

\begin{agdasnippet}
  \ExecuteMetaData[Generic/Examples/NbyE.tex]{nbepatterns}
\end{agdasnippet}

We finally define the algebra by case analysis: if the node at hand is an
application and its first component evaluates to a lambda, we can apply
the function to its argument using \AF{\_\$\$\_}. Otherwise we have either a
stuck application or a lambda, in other words we already have a value and can
simply return it using \AIC{C}.

\begin{agdasnippet}
  \ExecuteMetaData[Generic/Examples/NbyE.tex]{nbelc}
\end{agdasnippet}

We have not used the \AIC{⊥} constructor so \emph{if} the evaluation terminates
(by disabling totality checking we have lost all guarantees of the sort) we know
we will get a term in normal form. For instance, we can evaluate an untyped yet normalising
term {(λx. x) ((λx. x) (λx. x))} that normalises to {(λx. x)}:

\begin{agdasnippet}
  \ExecuteMetaData[Generic/Examples/NbyE.tex]{example}
\end{agdasnippet}

%% file: catalogue/unrolling.tex
\subsection{Binding as self-reference: representing cyclic structures}\label{def:colist}

Ghani, Hamana, Uustalu and Vene~\citeyear{ghani2006representing} have
demonstrated how Altenkirch and Reus' type-level de Bruijn
indices~\citeyear{altenkirch1999monadic} can be used to represent
potentially cyclic structures by a finite object. In their
representation each bound variable is a pointer to the node
that introduced it. Given that we are, at the top-level, only
interested in structures with no \scarequote{dangling pointers}, we introduce
the notation \AF{TM} \AB{d} to mean closed terms (i.e. terms of type
\AD{Tm} \AB{d} \AF{∞} \AIC{[]}).

A basic example of such a structure is a potentially cyclic list which
offers a choice of two constructors: \AIC{[]} which ends the list and
\AIC{\_::\_} which combines a head and a tail but also acts as a binder
for a self-reference; these pointers can be used by using the \AIC{var}
constructor which we have renamed \AIC{↶} (pronounced \literalquote{backpointer})
to match the domain-specific meaning.
We can see this approach in action in the examples
\AF{[0, 1]} and \AF{01↺} (pronounced \literalquote{0-1-cycle}) which describe
respectively a finite list containing
0 followed by 1 and a cyclic list starting with 0, then 1, and then
repeating the whole list again by referring to the first cons cell
represented here by the de Bruijn variable 1 (i.e. \AIC{s} \AIC{z}).

\noindent
\begin{minipage}{\textwidth}
  \begin{minipage}{0.55\textwidth}
    \ExecuteMetaData[Generic/Examples/Colist.tex]{clistD}
    \ExecuteMetaData[Generic/Examples/Colist.tex]{clistpat}
  \end{minipage}
  \begin{minipage}{0.35\textwidth}
    \ExecuteMetaData[Generic/Examples/Colist.tex]{zeroones}
  \end{minipage}
\end{minipage}

These finite representations are interesting in their own right
and we can use the generic semantics framework defined earlier
to manipulate them. A basic building block is the \AF{unroll}
function which takes a closed tree, exposes its top node and
unrolls any cycle which has it as its starting point. We can
decompose it using the \AF{plug} function which, given a closed
and an open term, closes the latter by plugging the former at
each free \AIC{`var} leaf. Noticing that \AF{plug}'s fundamental nature
is that of substituting a term for each leaf, it makes sense to
implement it by re-using the \AF{Substitution} semantics we already have.

\begin{agdasnippet}
  \ExecuteMetaData[Generic/Cofinite.tex]{plug}
  \ExecuteMetaData[Generic/Cofinite.tex]{unroll}
\end{agdasnippet}


However, one thing still out of our reach with our current tools is
the underlying cofinite trees these finite objects are meant to
represent. We start by defining the coinductive type corresponding to
them as the greatest fixpoint of a notion of layer. One layer of a
cofinite tree is precisely given by the meaning of its description
where we completely ignore the binding structure. We show with
\AF{01⋯} (mutually defined with \AF{10⋯}) the infinite list that
corresponds to the unfolding of the example \AF{01↺} given above.


\noindent
\begin{minipage}{\textwidth}
  \ExecuteMetaData[Generic/Cofinite.tex]{cotm}
  \newline
\begin{minipage}{0.5\textwidth}
  \ExecuteMetaData[Generic/Examples/Colist.tex]{zeroones2}
\end{minipage}
\begin{minipage}{0.49\textwidth}
  \ExecuteMetaData[Generic/Examples/Colist.tex]{zeroones3}
\end{minipage}
\end{minipage}

We can then make the connection between potentially cyclic
structures and the cofinite trees formal by giving an \AF{unfold}
function which, given a closed term, produces its unfolding.
The definition proceeds by unrolling the term's top layer and
co-recursively unfolding all the subterms.

\begin{agdasnippet}
 \ExecuteMetaData[Generic/Cofinite.tex]{unfold}
\end{agdasnippet}

Even if the
powerful notion of semantics described in Section~\ref{section:semantics}
cannot encompass all the traversals we may be interested in,
it provides us with reusable building blocks: the definition
of \AF{unfold} was made very simple by reusing the generic
program \AF{fmap} and the \AF{Substitution} semantics whilst
the definition of \AR{∞Tm} was made easy by reusing \AF{⟦\_⟧}.

%% file: catalogue/equality.tex
\subsection{Generic decidable equality for terms}

Haskell programmers are used to receiving help from the \codequote{deriving}
mechanism~\cite{DBLP:journals/entcs/HinzeJ00,DBLP:conf/haskell/MagalhaesDJL10}
to automatically generate common traversals for every inductive type
they define. Recalling that generic programming is normal programming
over a universe in a dependently typed
language~\cite{DBLP:conf/ifip2-1/AltenkirchM02}, we ought to be able to
deliver similar functionalities for syntaxes with binding.

We will focus in this section on the definition of an equality test. The
techniques used in this concrete example are general enough that they also
apply to the definition of an ordering test, a \texttt{Show} instance, etc.
In type theory we can do better than an uninformative boolean function
claiming that two terms are equal: we can implement a decision procedure
for propositional equality~\cite{DBLP:conf/icfp/LohM11} which either
returns a proof that its two inputs are equal or a proof that they
cannot possibly be.

The notion of decidability can be neatly formalised by an inductive family
with two constructors: a \AF{Set} \AB{P} is decidable if we can either say
\AIC{yes} and return a proof of \AB{P} or \AIC{no} and provide a proof of
the negation of \AB{P} (here, a proof that \AB{P} implies the empty type
\AD{⊥}).

\noindent
\begin{minipage}{\textwidth}
  \begin{minipage}[t]{0.45\textwidth}
    \ExecuteMetaData[Stdlib.tex]{bottom}
  \end{minipage}
  \begin{minipage}[t]{0.45\textwidth}
    \ExecuteMetaData[Stdlib.tex]{dec}
  \end{minipage}
\end{minipage}

\noindent To get acquainted with these new notions we can start by proving variable equality decidable.

\subsubsection{Deciding variable equality}

The type of the decision procedure for equality of variables is as follows:
given any two variables (of the same type, in the same context), the set of
equality proofs between them is \AD{Dec}idable.

\begin{agdasnippet}
\ExecuteMetaData[Generic/Equality.tex]{eqVarType}
\end{agdasnippet}

We can easily dismiss two trivial cases: if the two variables have distinct
head constructors then they cannot possibly be equal. Agda allows us to
dismiss the impossible premise of the function stored in the \AIC{no}
contructor by using an absurd pattern \AS{()}.

\begin{agdasnippet}
\ExecuteMetaData[Generic/Equality.tex]{eqVarNo}
\end{agdasnippet}

Otherwise if the two head constructors agree we can be in one of two
situations. If they are both \AIC{z} then we can conclude that the two
variables are indeed equal to each other.

\begin{agdasnippet}
\ExecuteMetaData[Generic/Equality.tex]{eqVarYesZ}
\end{agdasnippet}

Finally if the two variables are {(\AIC{s} \AB{v})} and {(\AIC{s} \AB{w})}
respectively then we need to check recursively whether \AB{v} is equal
to \AB{w}. If it is the case we can conclude by invoking the congruence
rule for \AIC{s}. If \AB{v} and \AB{w} are not equal then a proof that
{(\AIC{s} \AB{v})} and {(\AIC{s} \AB{w})} are will lead to a direct
contradiction by injectivity of the constructor \AIC{s}.

\begin{agdasnippet}
\ExecuteMetaData[Generic/Equality.tex]{eqVarYesS}
\end{agdasnippet}

\subsubsection{Deciding term equality}

The constructor \AIC{`σ} for descriptions gives us the ability to store
values of any \AF{Set} in terms. For some of these \AF{Set}s (e.g.
{(\AD{ℕ} → \AD{ℕ})}), equality is not decidable. As a consequence
our decision procedure will be conditioned to the satisfaction of a
certain set of \AF{Constraints} which we can compute from the \AD{Desc}
itself. We demand that we are
able to decide equality for all of the \AF{Set}s mentioned in a description.

\begin{agdasnippet}
\ExecuteMetaData[Generic/Equality.tex]{constraints}
\end{agdasnippet}

Remembering that our descriptions are given a semantics as a big right-nested
product terminated by an equality constraint, we realise that proving decidable
equality will entail proving equality between proofs of equality. We are happy
to assume Streicher's axiom K~\cite{DBLP:conf/lics/HofmannS94} to easily
dismiss this case. A more conservative approach would be to demand that equality
is decidable on the index type \AB{I} and to then use the classic Hedberg
construction~\cite{DBLP:journals/jfp/Hedberg98} to recover uniqueness of
identity proofs for \AB{I}.

Assuming that the constraints computed by {(\AF{Constraints} \AB{d})} are
satisfied, we define the decision procedure for equality of terms together
with its equivalent for bodies. The function \AF{eq\textasciicircum{}Tm}
is a straightforward case analysis dismissing trivially impossible cases
where terms have distinct head constructors (\AIC{`var} vs. \AIC{`con})
and using either \AF{eq\textasciicircum{}Var} or \AF{eq\textasciicircum{}⟦⟧}
otherwise. The latter is defined by induction over \AB{e}. The somewhat
verbose definitions are not enlightening so we leave them out here.

\begin{agdasnippet}
\ExecuteMetaData[Generic/Equality.tex]{eqTmType}
\end{agdasnippet}

We now have an informative decision procedure for equality between terms
provided that the syntax they belong to satisfies a set of constraints.
Other generic functions and decision procedures can be defined
following the same approach: implement a similar function for variables
first, compute a set of constraints, and demonstrate that they are
sufficient to handle any input term.